\documentclass[pre,aps,twocolumn,showpacs,groupedaddress,superscriptaddress,amsmath,amssymb,10pt]{revtex4-1}
\pdfoutput=1
\usepackage{graphicx}
\usepackage{dcolumn}
\usepackage{bm}
\usepackage{epsfig,amsmath}
\usepackage{subfigure}

\begin{document}

\title{Understanding and Controlling Regime Switching in Molecular Diffusion}

\author{S.~Hallerberg}
\affiliation{Network Dynamics, Max Planck Institute for Dynamics and Self-Organization (MPIDS), 37077 G{\"o}ttingen, Germany}
\affiliation{Institute of Physics, TU Chemnitz, 09107 Chemnitz, Germany}

\author{A.~S.~de Wijn}
\affiliation{Department of Physics, Stockholm University, 106 91 Stockholm, Sweden}
\affiliation{Radboud University Nijmegen, Institute for Molecules and Materials, Heyendaalseweg 135, 6525AJ Nijmegen, The Netherlands}

 \begin{abstract}
 Diffusion can be strongly affected by ballistic flights (long jumps) as well as long-lived sticking trajectories (long sticks).
 Using statistical inference techniques in the spirit of Granger causality, 
 we investigate the appearance of long jumps and sticks in molecular-dynamics simulations of diffusion in a prototype system, a benzene molecule on a graphite substrate.
 We find that specific fluctuations in certain, but not all, internal degrees of freedom of the molecule can be linked to either long jumps or sticks.
 Furthermore, by changing the prevalence of these predictors with an outside influence, the diffusion of the molecule can be controlled.
 The approach presented in this proof of concept study is very generic, and can
 be applied to larger and more complex molecules. 
 Additionally, the predictor variables can be chosen in a general way so as to be accessible in experiments, making the method feasible for control of diffusion in applications.
 Our results also demonstrate that data-mining techniques can be used to investigate the phase-space structure of high-dimensional nonlinear dynamical systems.
 %
 
 \end{abstract}

\maketitle
\section{Introduction}
The diffusion of molecules and clusters of atoms on substrates is of substantial importance for the operation of 
nano-scale devices, control of chemical reactions, catalysis, and self-assembly.
Experiments~\cite{longjumps,grazynajumps1} as well as numerical simulations~\cite{braun, maruyamagold, sonnleitner} have revealed that {\it long jumps}, i.e. long-lived ballistic trajectories can strongly affect the surface diffusion of single atoms, molecules, and nano-scale clusters.
Such movements, named flights in the dynamical systems community, have been studied for the motion of point particles in periodic lattices \cite{Geiselelectrons}, as well as on much larger scales such as the geographic spread of diseases~\cite{largescaleflights}.
Apart from long jumps, similar systems can also exhibit the opposite behavior: staying in a vicinity for an inordinately long amount of time.
These kinds of jumps and sticks can result in anomalous diffusion, i.e. diffusion with a mean square displacement that grows faster or slower than linearly with time.
Even when the anomalousness is destroyed by noise or some other mechanism~\cite{sokolovpre2004}, jumps or sticks may remain and strongly affect the diffusion.

The diffusion of larger molecules is affected by the dynamics of their internal degrees of freedom, which form an energy reservoir capable of absorbing and releasing kinetic energy~\cite{onswdps13,astridbenzene}.
The overall diffusion of these molecules is normal, due to the strongly chaotic internal degrees of freedom of the molecule~\cite{onswdps13,astridbenzene} and the thermal noise from the substrate, in contrast to the anomalous diffusion of point particles in periodic lattices \cite{Geiselelectrons,sokolovpre2004,sokolovprl2004}.
Nevertheless, the molecule's trajectory contains sections in which it temporarily behaves similarly to an anomalously diffusing object, moving ballistically (long jumps) or remaining close to the vicinity of one unit cell (long sticks).

Diffusion in dynamical systems has also been observed in some cases to switch between long-lived movements and normal diffusive behavior~\cite{eduardo,gilbert}.
%
%
Many methods exist for studying dynamical systems, but almost all of them have been developed for simplified low-dimensional systems with typically one- to four-dimensional phase spaces (see, e.g.~\cite{rainerbook}).  
For dynamical systems of higher dimension, few useful approaches exist.
In this article we propose a data-mining approach, to reveal links between energy fluctuations in the internal degrees of freedom of a high-dimensional dynamical system (benzene diffusing on graphite) on the one hand and the rare events (long jumps and sticks) in the diffusion on the other.
\begin{figure*}
\begin{minipage}[t!!!]{0.35\textwidth}
\epsfig{figure=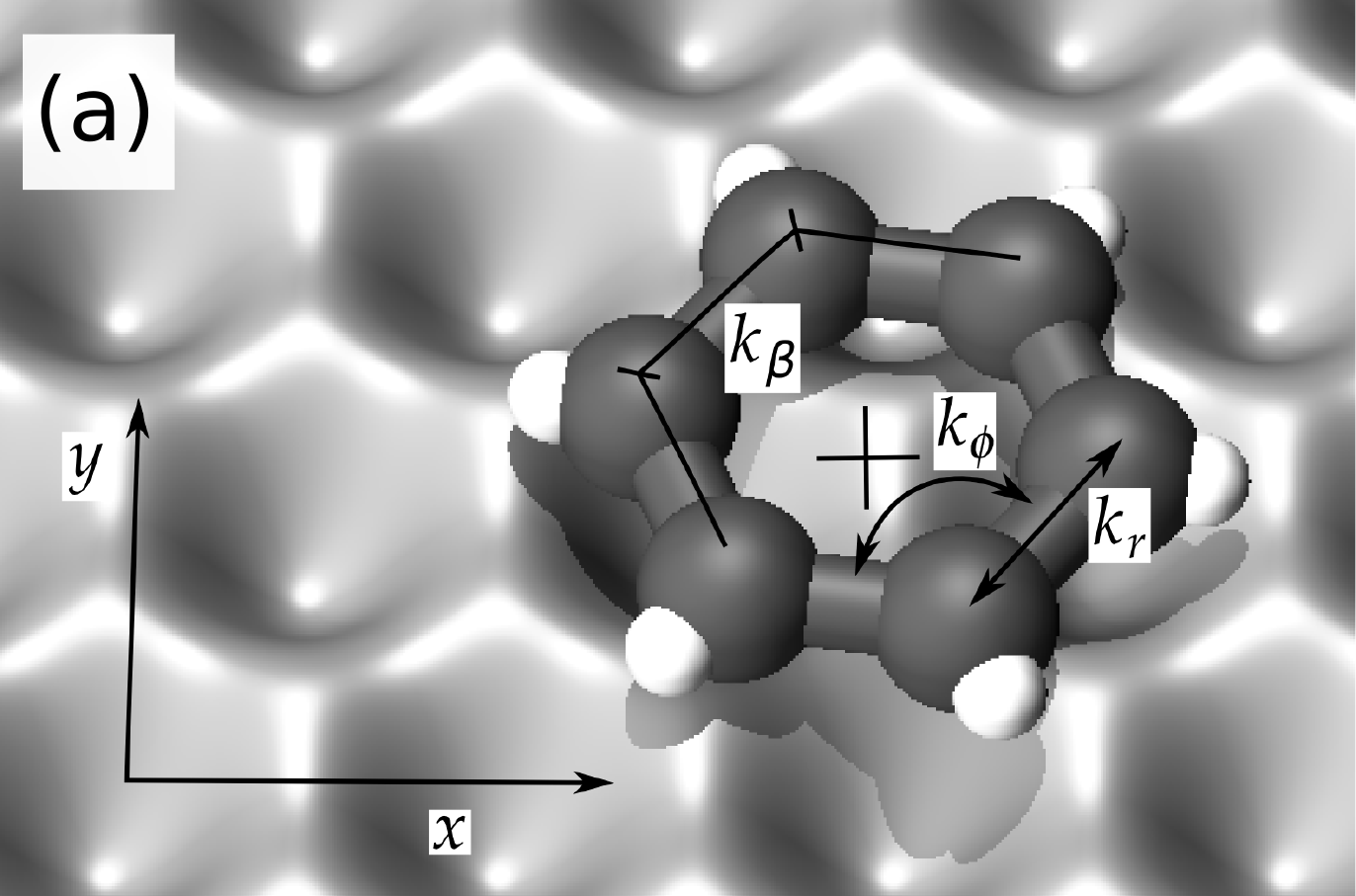,width=1.0\textwidth}
\end{minipage}
\hspace{0.7cm}
\begin{minipage}[t!!!]{0.60\textwidth}
\epsfig{figure=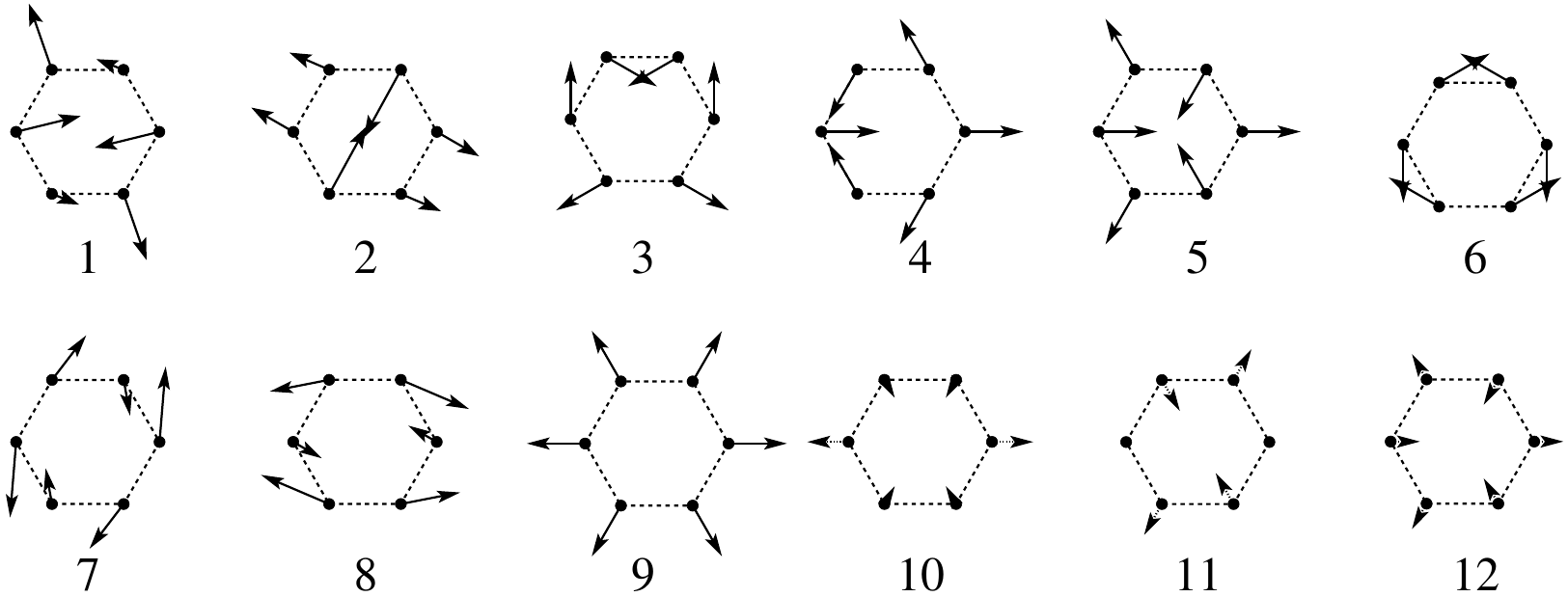,width=1.0\textwidth}
\put(-320, 105){\large \sf(b)}
\end{minipage}
\caption{\label{fig:eigenmodes} (a) The prototype system, a benzene molecule on a graphite substrate~\cite{onswdps13,astridbenzene}.
The internal dynamics consist of bond stretching, bending, and torsion.
(b) The 12 vibrational eigenmodes of the linearized system (numbered arbitrarily). 
The arrows indicate the directions of the vibrations.  For modes 10, 11, and 12, which are torsion modes, the vibrations are out-of-plane.  The motion for these modes (away from the reader or towards them) is indicated with respect to the center of the hexagon.
}
\end{figure*}
Links between two variables or events are often studied using averaged quantities, such as cross-correlation functions, mutual information \cite{cover}, Kullback-Leibler divergences \cite{KL1951}, and tests for Granger causality \cite{Granger}.
However, these approaches fail when the events under study are rare and their contribution to the average is negligible.
Therefore we use a conceptually different approach, i.e., we use statistical-inference techniques and analyze the success rate of the inference using Receiver Operator Characteristic curves (ROC-curves) \cite{Egans}, which are a common measure for the success of classification algorithms in machine learning and data mining \cite{Physa, Sarah3, BogachevKireenkovNifontovBunde2009, BogachevBunde-EPL-2009}.
Using these prediction methods in order to identify links between variables and future discrete events provides a simplified framework for testing for Granger causality in point processes.
A conceptually similar approach has recently been studied in the context of neuro-science \cite{Kim2011}.

Having identified relevant predictors, we manipulate the diffusion of the simulated molecule by deliberately triggering them.
Our approach is very general and can easily be extended to the design of mechanisms that alter the diffusion of larger molecules on other substrates.
%
%
This paper thus presents a proof-of-concept study, demonstrating that data-mining techniques can be used to extract useful information from molecular-dynamics simulations.
The arrows indicate the amplitude and direction of the atomic motion, being within (mode 1 - 9) or orthogonal to the plane of the molecule (torsion modes 10, 11, and 12).
Some modes are degenerate in sets of two, namely (1, 2) , (3, 4), (7, 8), and (10, 11).
\section{Modelling a Benzene Molecule diffusing on Graphite}
As a prototype system for diffusion of large molecules, we consider a benzene molecule on a graphite substrate [Fig.~\ref{fig:eigenmodes}(a)].
A particularly suitable model for investigating the dynamical properties of this system was developed in Ref.~\cite{onswdps13}.
It contains the essential nonlinear dynamics, without including any of the myriad of extra complications that are not of interest here.
Similar models have been successfully used for more complicated, less symmetric molecules~\cite{cppyrrolethiophene}.
The dynamics are described with a classical atomistic force field, based on the Tripos 5.2 force field~\cite{tripos5.2}.
The hydrogen atoms are treated in a mean-field approximation, as their dynamics cannot be described reliably classically.

Let $\vec{r}_i$
denote the position
of the $i$-th CH complex, ordered in such a way that $i$ and $(i+1\mod 6)$ are neighbors in the benzene ring.
Let $\phi_i$ and $\beta_i$ be the angles between the bonds and the torsion angles respectively.
The internal potential energy of the benzene molecule is written as
a sum over bending, stretching, and torsion of the bonds between the carbon atoms,
\begin{eqnarray}
V_\mathrm{molecule} (\vec{r}_1,\ldots,\vec{r}_6) &=&
{\textstyle \frac12} k_r \sum_{i=1}^6 (\|\vec{r}_{(i+1)(\mathrm{mod}~6)}-\vec{r}_{i}\|-r_0)^2
\nonumber\\
&&\null
+ {\textstyle \frac12} k_\phi \sum_{i=1}^6 (\phi_i-{\textstyle \frac23}\pi)^2
\nonumber\\
&&\null
+ k_\beta \sum_{i=1}^6
[1+\cos (2 \beta_i)]~,
\label{eq:Vbenzene}
\end{eqnarray}
where $k_r$ and $r_0$ are the C-C stretching force constant and equilibrium distance, while $k_\phi$ and $k_\beta$ are the effective bending force constant and the effective torsion constant.
%
In this work, we use the same values as in Refs.~\cite{onswdps13,astridbenzene}, $r_0 = 1.47$~\AA,  $k_r = 60.7~\mathrm{eV}/$\AA$^2$, $k_\phi = 6.85~\mathrm{eV}/\mathrm{rad}^2$, and $k_\beta = 0.247~\mathrm{eV}$.
The internal degrees of freedom of the model molecule display chaotic dynamics~\cite{onswdps13}. 
%
This acts as effective noise and influences the friction and diffusion of the molecule on the substrate~\cite{astridbenzene}.  

As in Ref.~\cite{astridbenzene}, we represent the substrate using a three-di\-men\-sional substrate potential for each CH complex that is composed of a two-dimensional hexagonal sinusoidal potential in the $xy$ plane and a harmonic term in the $z$ direction,
\begin{eqnarray}
V_\mathrm{CH}(\vec{r}) = -\frac{2 V_\mathrm{c}}{9}\left[2\cos\left(\frac{2 \pi x}{a \sqrt{3}}\right)\cos\left(\frac{2 \pi y }{3 a}\right)
\right.\nonumber\\
\left.\null + \cos\left(\frac{4 \pi y }{3 a}\right)\right] + V_c \frac{8\pi^2}{27a^2} z^2~,
\label{eq:Vsubstrate}
\end{eqnarray}
where $V_\mathrm{c}= 25$~meV is the potential corrugation, and $a=1.42$~\AA\ is the in-layer inter-atomic distance of graphite.
A Langevin thermostat with temperature $293$~K and damping parameter of 0.0025/ps is applied to each CH complex.
It model the thermal fluctuations and damping due to the heatbath of the substrate.
The viscous damping parameter has been chosen sufficiently low for the diffusion to be dominated by long jumps and sticks.

Realistic damping parameters for small molecules are typically higher, around 1/ps.  For larger molecules, little information is available.  
It is known, however, that for larger interfaces, friction can become extremely low due to structural incompatibility~\cite{shinjo}.
As long jumps have been observed in experiments on large molecules~\cite{longjumps}, we know that in some cases the damping is in the regime that allows jumps to occur.
An example of a trajectory with long jumps is shown in Fig.~\ref{fig:trajectories}.
%
A total of 16 molecular-dynamics simulations were run for a time of 1.2~$\mu$s each.
The $16$ simulations differ only in their randomly chosen initial conditions and the precise realization of the applied Langevin thermostat.
The coordinates of the center of mass and configuration of the internal degrees of freedom were stored every $\Delta t= 0.24$~ps.  
%

%
%
\section{Identifying anomalous movements}
\begin{figure*}[t!!!]
\vskip-1\bigskipamount
\centerline{
\hskip-2.5\bigskipamount
\begin{minipage}[t!!!]{0.32\textwidth}
\epsfig{figure=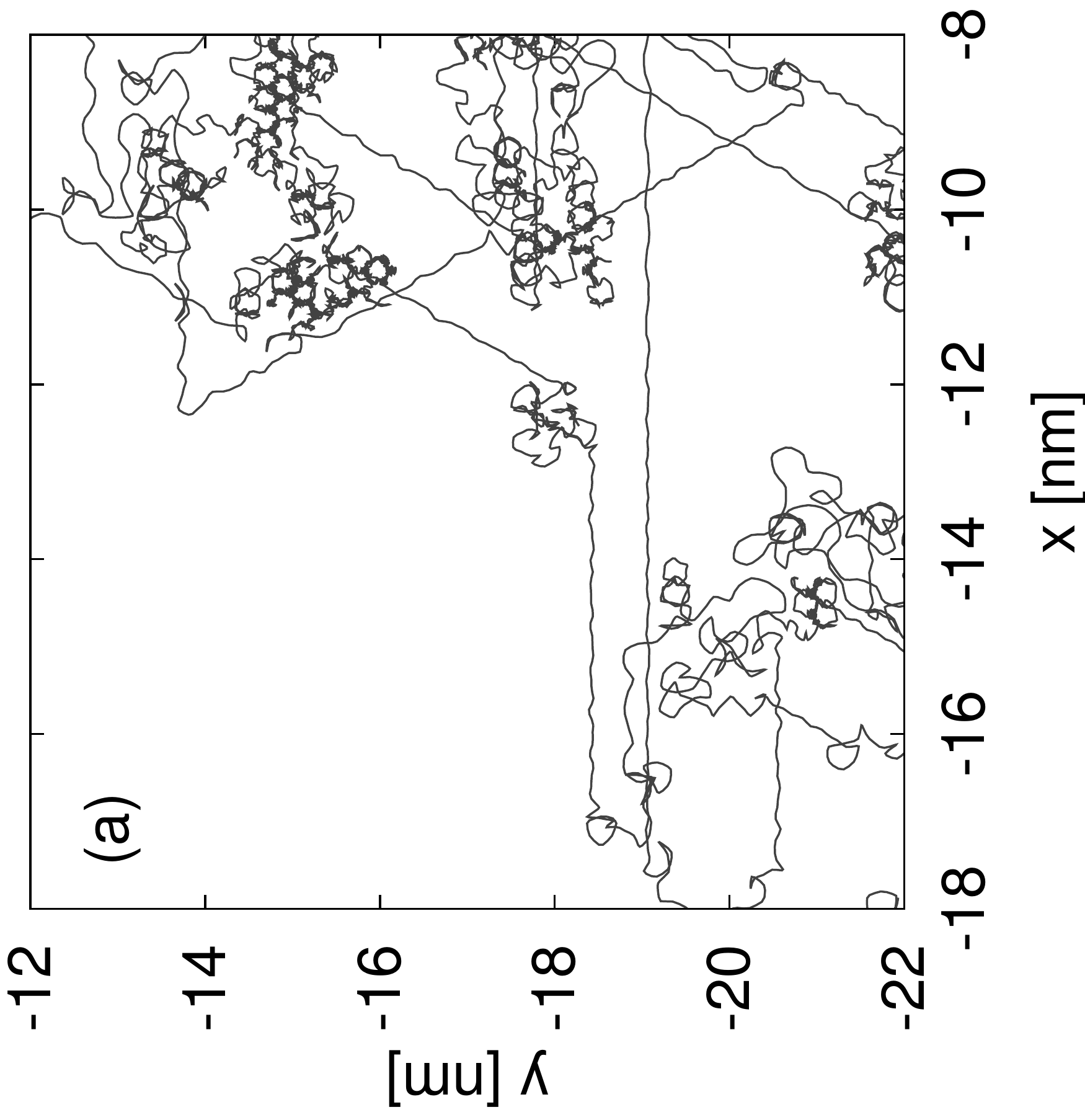,  angle=-90, width=1.06\textwidth}
\end{minipage}
\hspace{0.2cm}
\begin{minipage}[t!!!]{0.32\textwidth}
\epsfig{figure=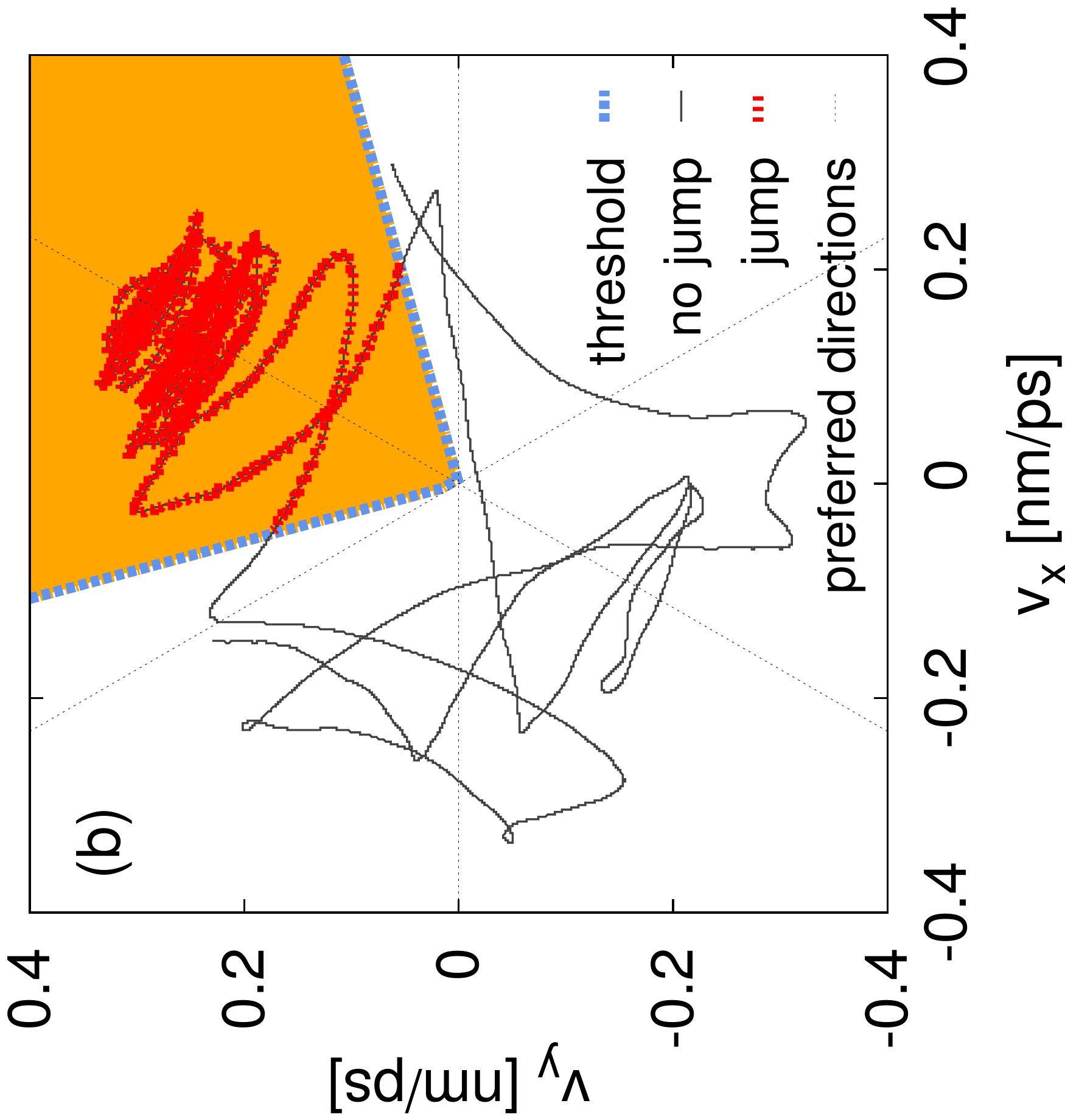, angle=-90, width=1.06\textwidth}
\end{minipage}
\hspace{-0.6cm}
\begin{minipage}[t!!!]{0.32\textwidth}
\epsfig{figure=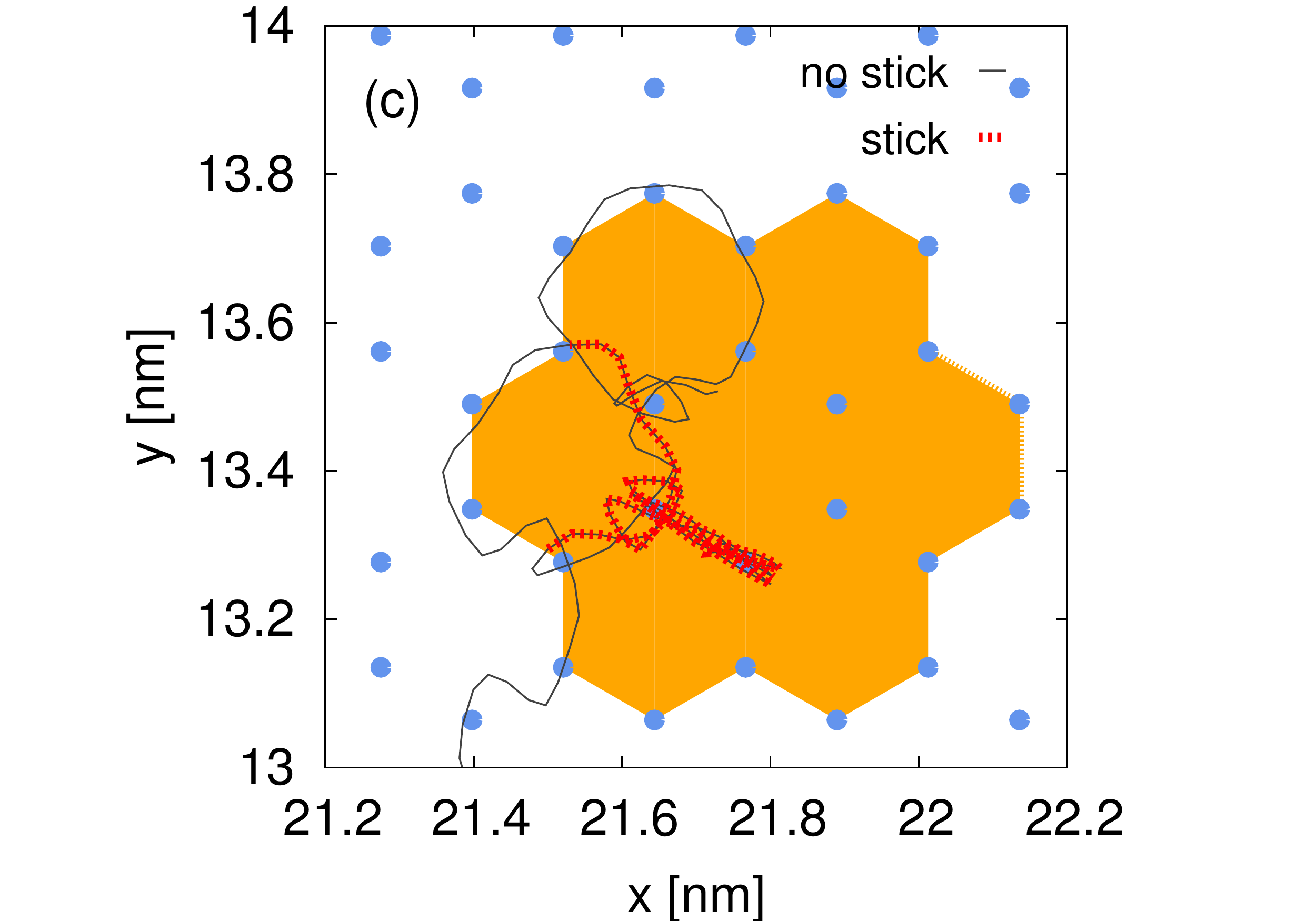, angle=-90, width=1.43\textwidth}
\end{minipage}
}
\vskip-0.5\bigskipamount
\caption{\label{fig:trajectories}
(Color online) Detecting ballistic flights and subdiffusive sticks.
(a) Example of trajectories of a simulated benzene molecule on a graphite substrate in real space including ballistic flights (long jumps) and subdiffusive sticks. 
In between long jumps the mean-square displacement of the center of mass of the molecule grows linearly with time and diffusion is much slower.
(b) A short section of the trajectory in velocity space with the long jump highlighted. 
(c) A short section of the trajectory in real space with a stick highlighted.
}
\end{figure*}
We define a section of the trajectory as a long jump if the direction of the velocity vector remains within the vicinity of one orientation for $\tau$ time steps of length $\Delta t$.  
As the ballistic flights follow the substrate geometry, we detect jumps using angular sectors of $[c 60^\circ-45^\circ, c 60^\circ+45^\circ]$, with $c=0,1,2,3,4,5$ [e.g., the shaded area in Fig.~\ref{fig:trajectories}(b)].
Similarly, a section of the trajectory is taken to be a long stick if the molecule stays within a roughly hexagonal neighborhood of seven hexagons of carbon atoms on the substrate.
This is shown in Fig.~\ref{fig:trajectories}(c).

\begin{figure}
\vskip-1.0\bigskipamount
\hskip0.2\bigskipamount
\centerline{
\includegraphics[width=0.5\textwidth]{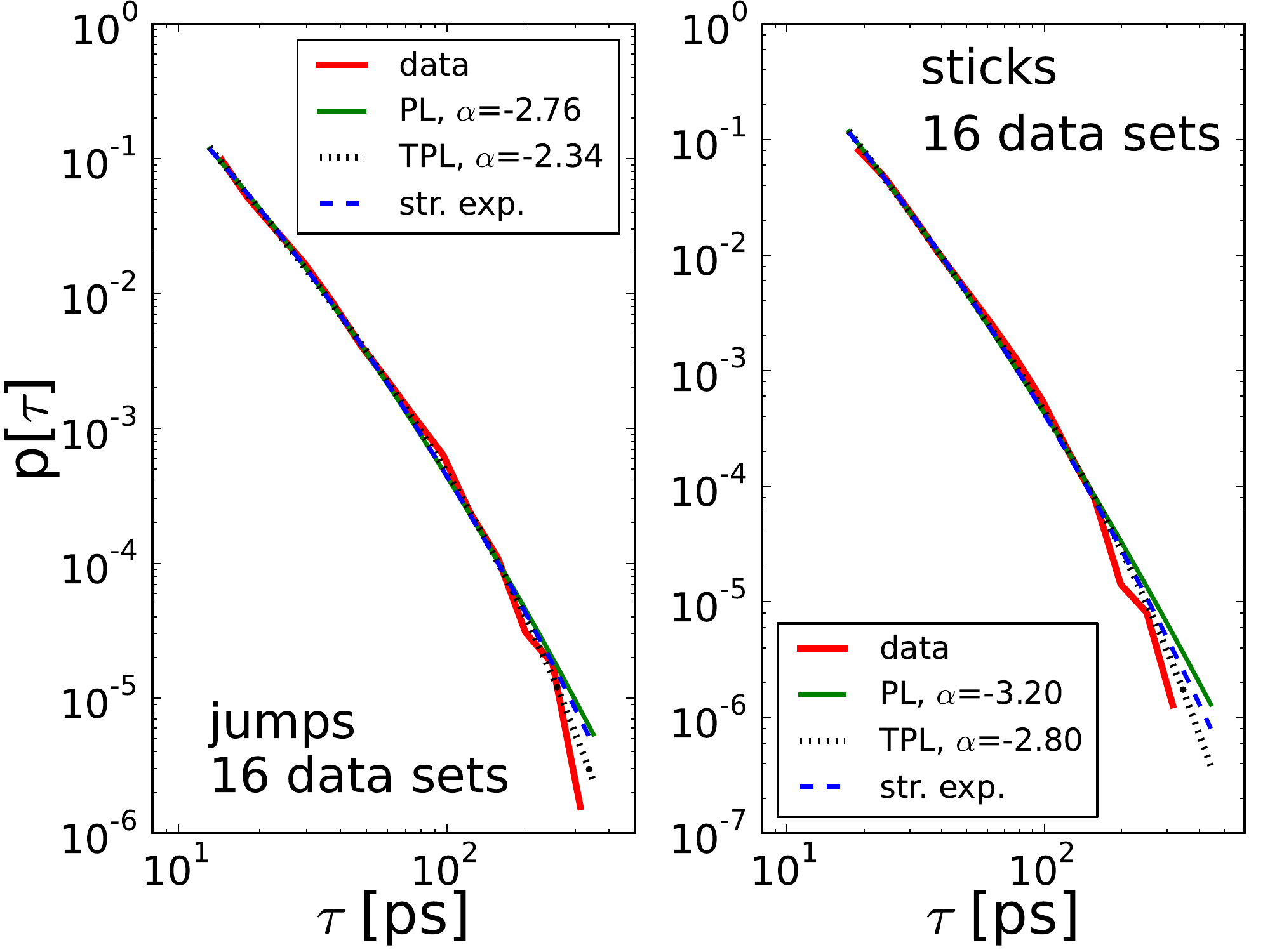}
}
\vskip-0.5\bigskipamount
\caption{\label{fig:jumpstickdist} (Color online) Estimating the density of jump and stick time length distributions by fitting several heavy-tailed distributions. A loglikelihood ratio test revealed that the truncated power law (TPL) is the most appropriate fit among the distributions tested [power law (PL), truncated power law (TPL), stretched exponential, lognormal]. Note that the difference between the PL fit and the lognormal fit are not visible. Consequently only the PL fit is labeled.}
\end{figure}

The distributions of the durations of jumps and sticks are shown in Fig.~\ref{fig:jumpstickdist}.
To estimate and fit these distributions, we use data of the time lengths of jumps and sticks as estimated from each of the $16$ simulations, as well as two concatenated data sets that contain jump or stick time lengths from all $16$ simulations.
We follow the suggestions of \cite{Shalizi} concerning appropriate ways of fitting heavy-tailed distributions and especially power laws (PLs)
\begin{equation}
\rho(\tau) = \frac{\alpha -1}{\tau_\mathrm{min}} \left(\frac{\tau}{\tau_\mathrm{min}}\right)^{\alpha}\quad,
\end{equation}
with $\tau_\mathrm{min}$ being a lower cutoff values.
 We apply the software package {\sl power law} \cite{alstott2013} to estimate $\alpha$ using a maximum likelihood estimator and adapting the minimum $\tau_\mathrm{min}$ such that the resulting density minimizes the Kolmogorov-Smirnov distance.
We compare maximum likelihood fits of a power law (PL), a truncated power law (TPL), a lognormal distribution, and a stretched exponential.
Among these distributions, we find that jumps and sticks are both best described by truncated power laws, with exponents $-2.45$ (long jumps) and $-2.87$ (sticks).
The upper limit of the power law scaling is likely due to the exponential decay of correlations on long time scales of order $1/\eta=$~400~ps, enforced by the Langevin thermostat used in the simulations.
In Fig.~\ref{fig:jumpstickdist} this exponential decay on larger time scales is visible as the small deviation between the tails of the fitted power-laws and the exponential tails of the fitted truncated power law.
\begin{figure*}[t!!!]
\vskip-2.5\bigskipamount
\hskip1.5\bigskipamount
\centerline{
\epsfig{figure=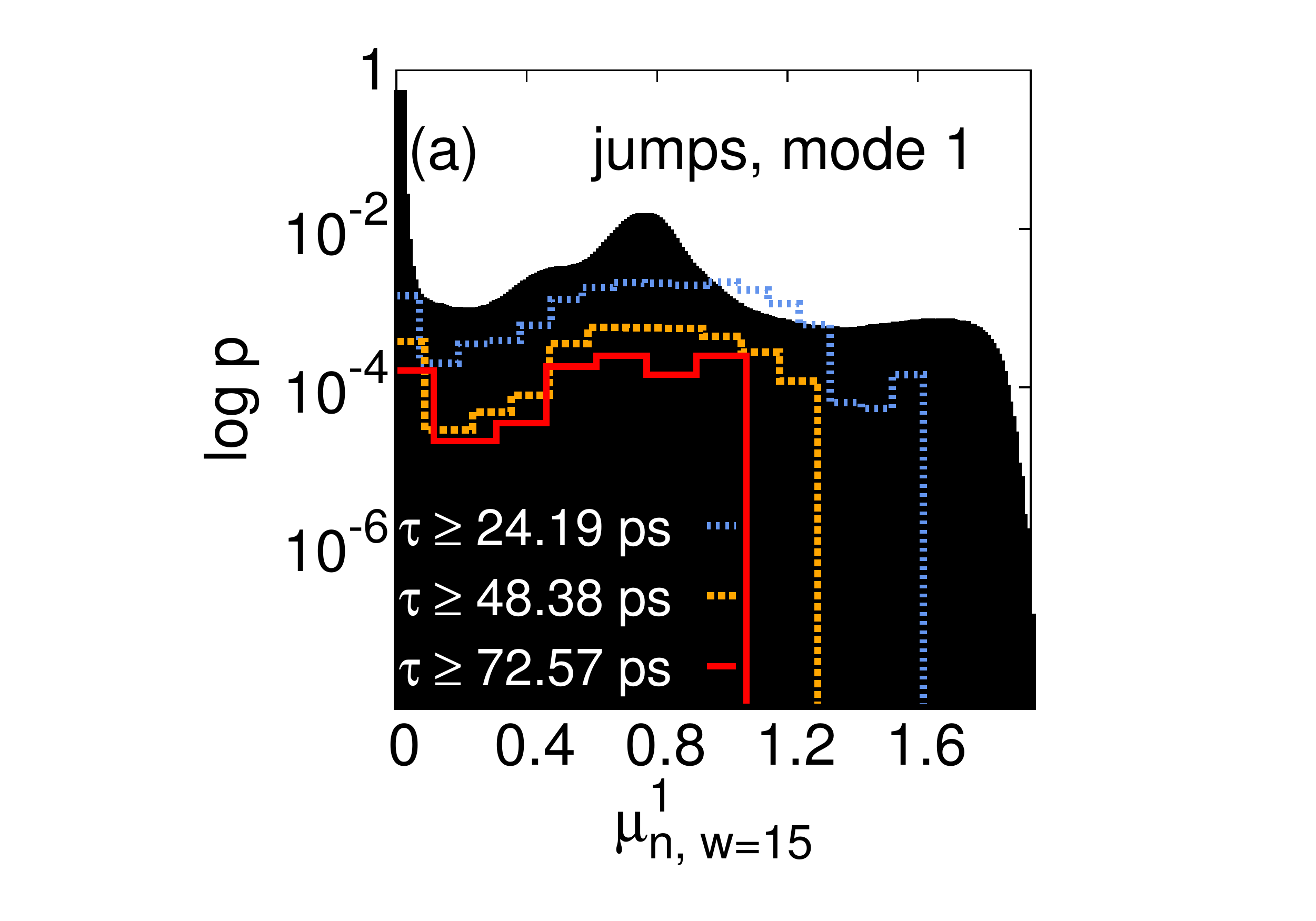,angle=-90,width=7cm}
\hspace{-2.7cm}
\epsfig{figure=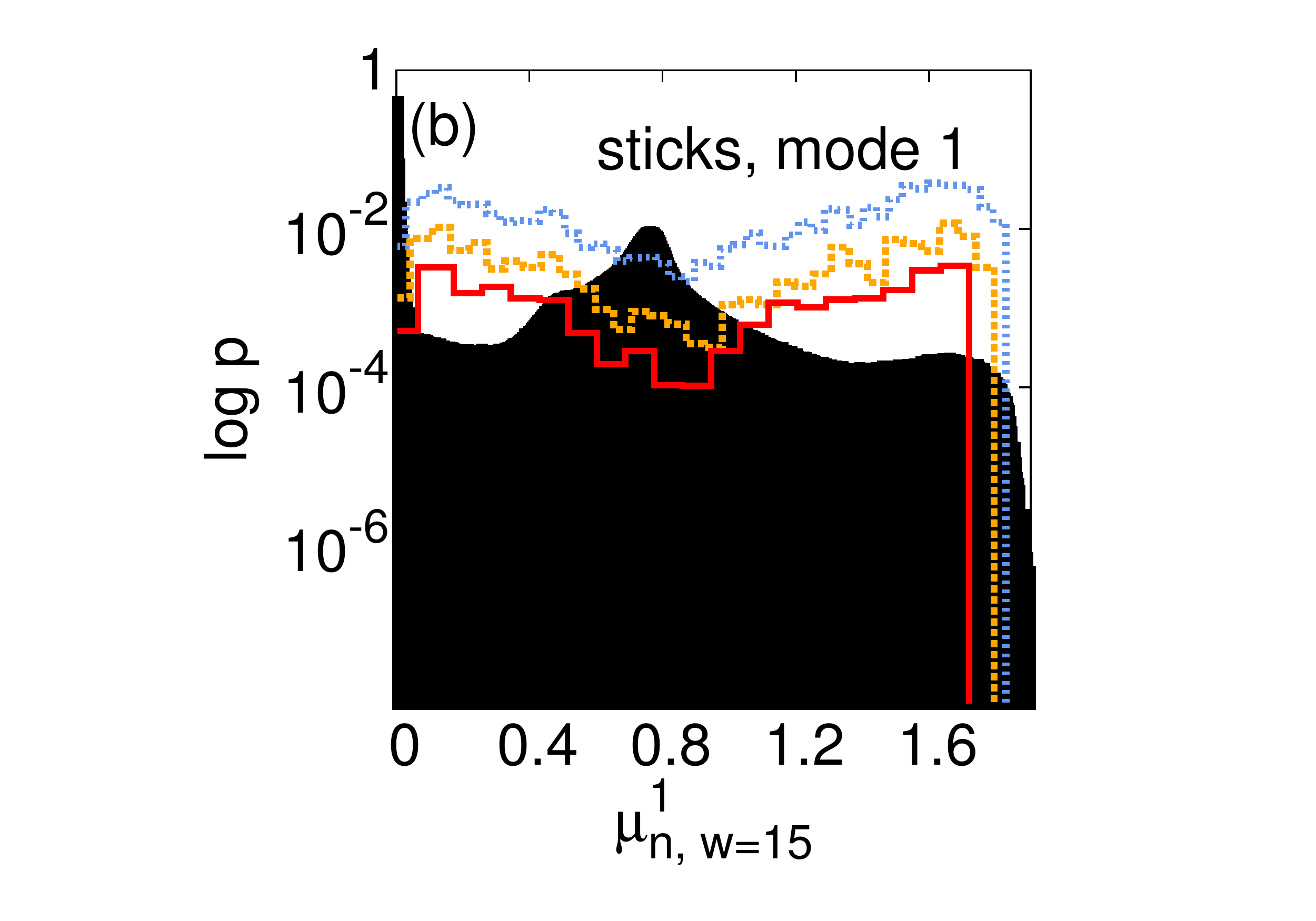,angle=-90,width=7cm}
\hspace{-2.7cm}
\epsfig{figure=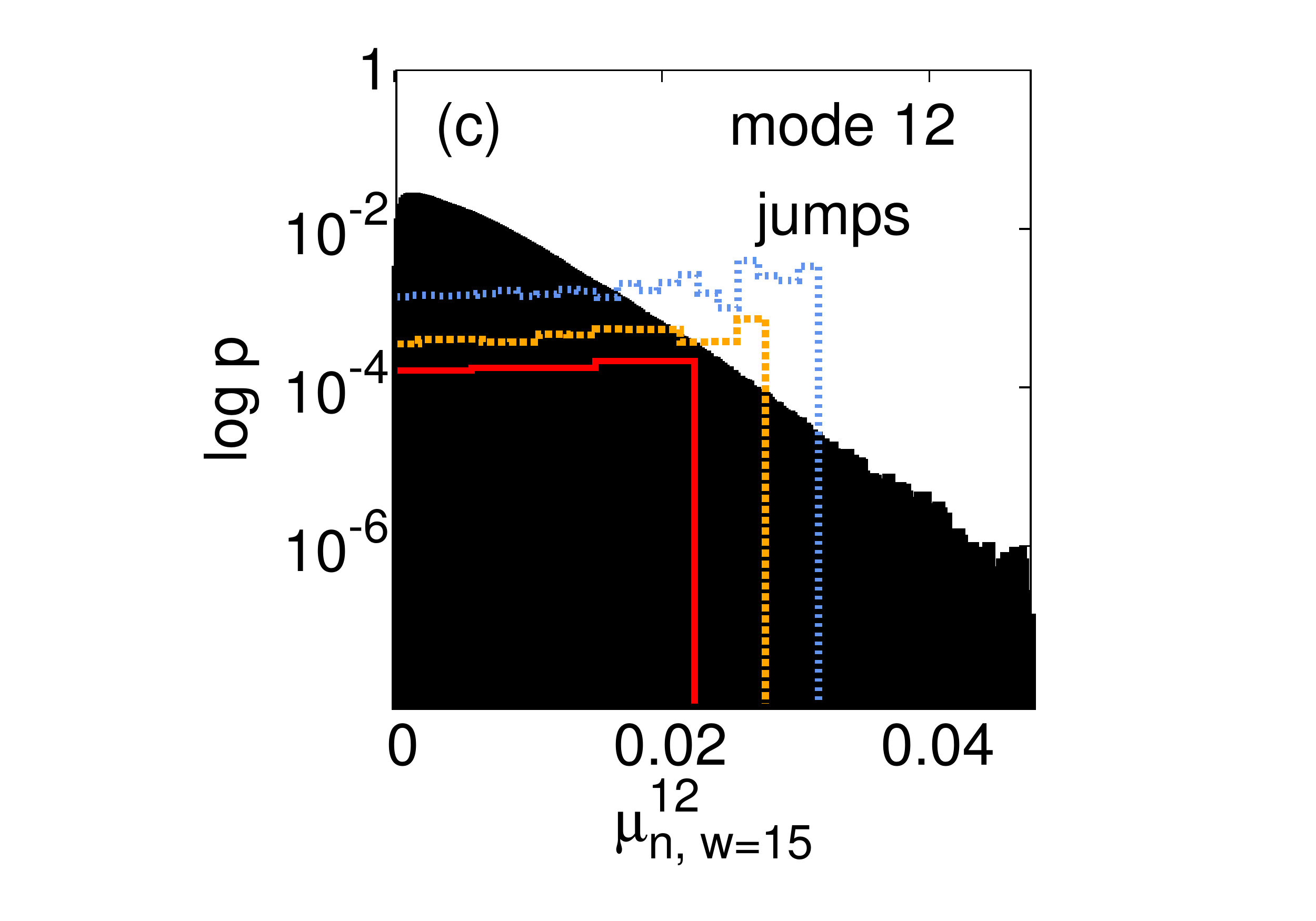, angle=-90,width=7cm}
\hspace{-2.7cm}
\epsfig{figure=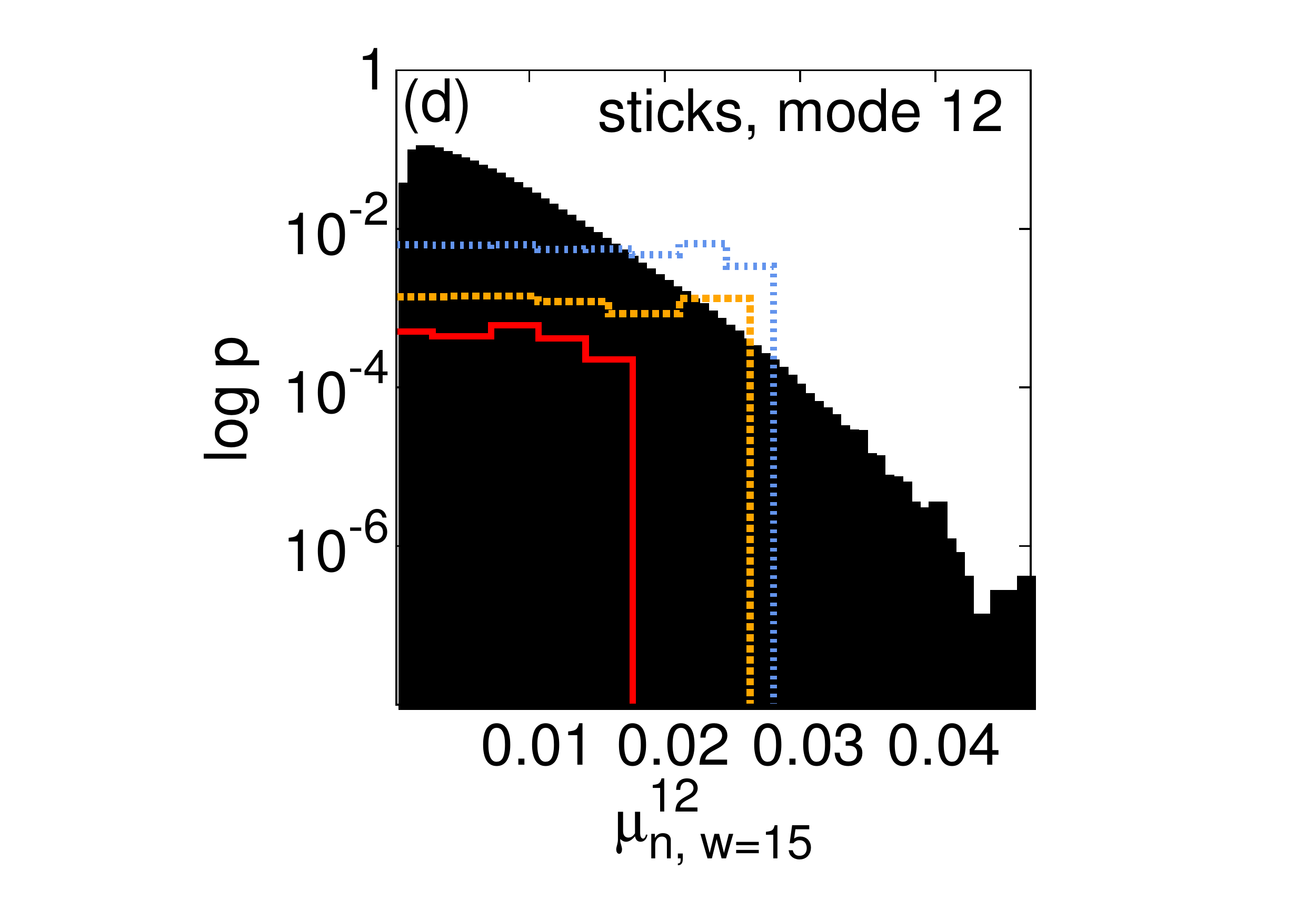, angle=-90,width=7cm}
}
\vskip-1.2\bigskipamount
\centerline{
\epsfig{figure=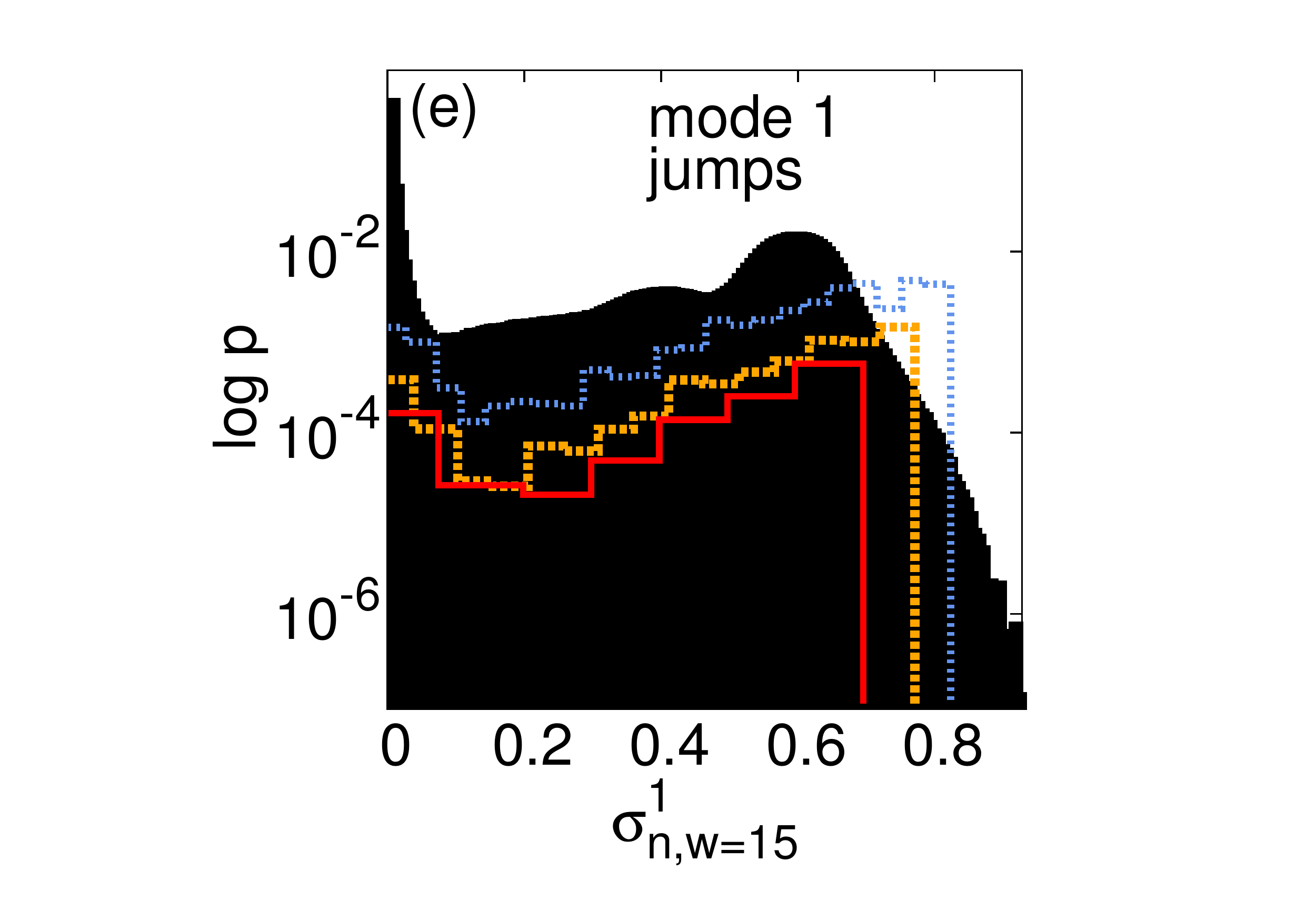,angle=-90,width=7cm}
\hspace*{-2.7cm}
\epsfig{figure=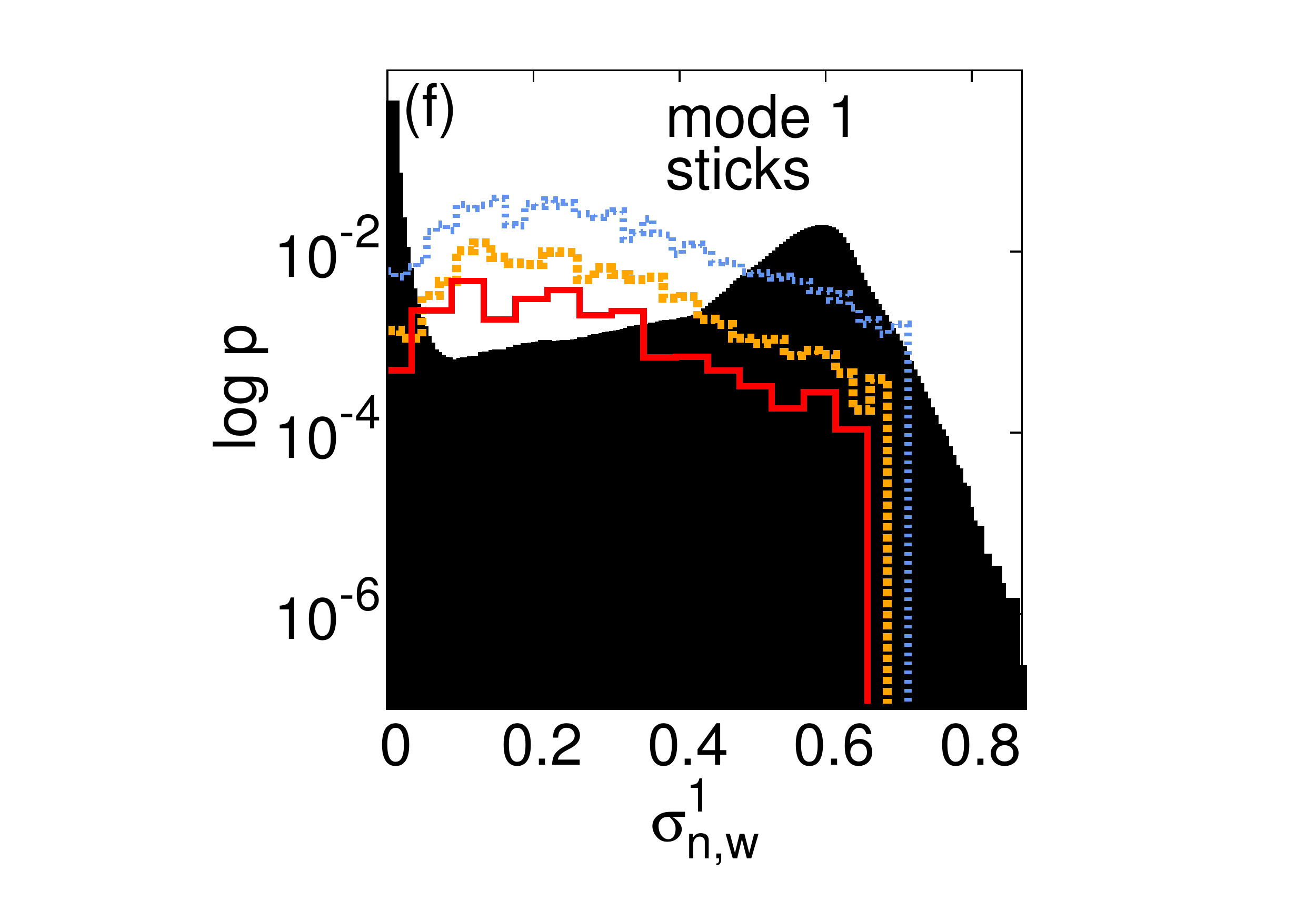,angle=-90,width=7cm}
\hspace{-2.7cm}
\epsfig{figure=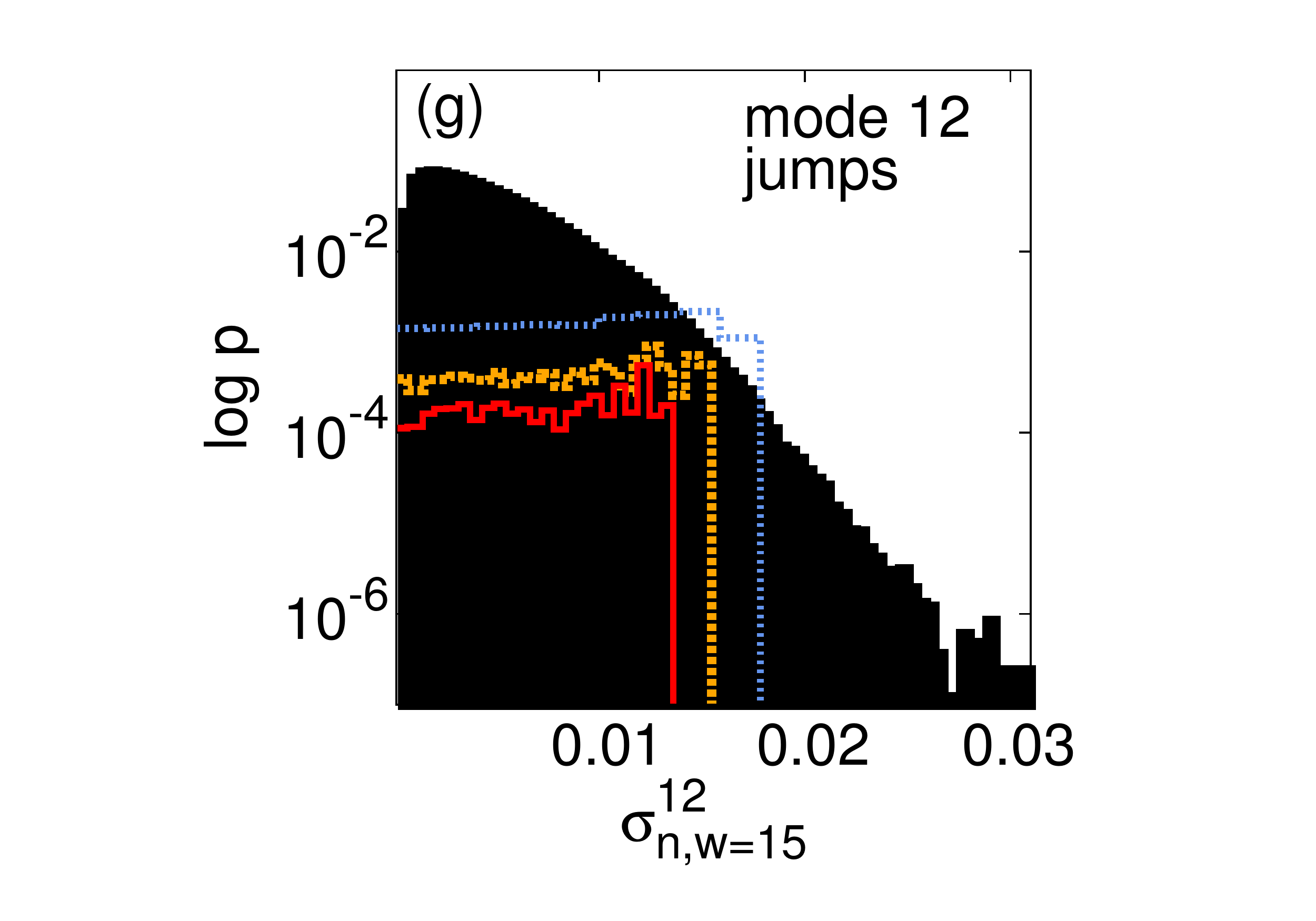, angle=-90,width=7cm}
\hspace{-2.7cm}
\epsfig{figure=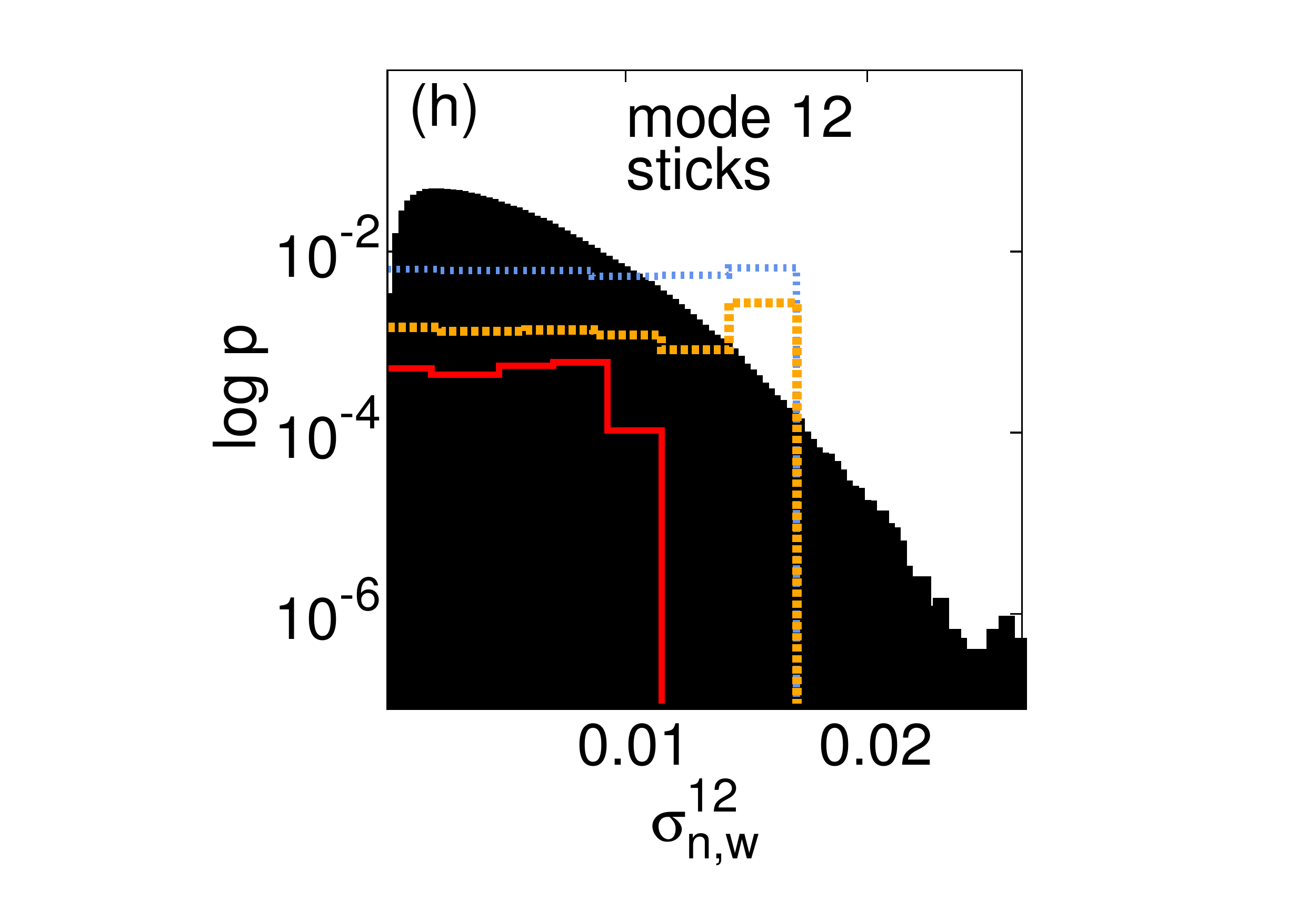, angle=-90,width=7cm}
}
\caption{\label{fig:cpdfs} (Color online) Examples of predictor distributions for modes that are linked (a), (b), (e), (f) or not linked (others) to the occurrence of sticks.
Histograms of CPDFs $p(s_n \geq \tau |y_n)$ are plotted as lines over the marginal probability distribution $p(y_n)$ (plotted as black bars). }
\end{figure*}
We also calculate the variance of the estimated exponent $\alpha$ among the ensemble of $16$ sets of simulation data.
The precise value of $\alpha$ varies among the different runs.
However, most of the values are within an 0.99 confidence interval centered around the ensemble means, which are $\alpha_\mathrm{jump}=-2.45$ for jumps and $\alpha_\mathrm{sticks}=-2.78$ for sticks. 
These means are also close to the values estimated using the concatenated data set of all $16$ simulations (see Fig.~\ref{fig:jumpstickdist}).
%
%
\section{Predicting long jumps and sticks}
By considering the results of the simulations as a time series, we can search for structures that precede or coincide with long jumps or sticks.
As the full trajectory of every CH complex in the simulated molecule is known, any function of the coordinates can, in principle, be used as an indicatory variable $y_n$.
We are therefore free to choose physically relevant quantities that could be influenced in experiments, namely the energy stored in the vibrational modes.
These energies we approximate by linearizing the Hamiltonian around the equilibrium solution.
Since the system has 18 degrees of freedom, of which 3 are center-of-mass translation and 3 are rotation, there are 12 eigenvibrations, shown in Fig.~\ref{fig:eigenmodes}.
%
The 36-dimensional phase space is thus summarized by the energies stored in these vibrations.
The 12 energies $x_n^{i}$ ($i=1, 2, ..., 12$) are recorded as a multivariate time series $\{ \mathbf{x}_n \} = \{ (x_n^{1},x_n^{2}, ...,  x_n^{12}) \}$, at discrete time instances $t = t_0 + n \Delta t$.
As predictors $y_n \in (\mu^{1}_{n}, ... , \mu^{12}_{n}, \sigma^{1}_{n} ..., \sigma^{12}_{n})$, we consider sliding window averages $\mu_n^{i}= w^{-1}\sum_{l=n-w}^n x^i_l$ and sliding window estimates of the standard deviations $\sigma_n^{i} =(w-1)^{-1}\sum_{l=n-w}^n (x^i_l - \mu^{i}_{n})^{2}$, ($i=1, 2, ..., 12$). 
The sliding window was chosen to start $w$ steps $\Delta t$ before the time step $n$ in which the prediction of the event occurring at time $n+l$ is made.
The values of $w$ shown here are $w=15$ for long jumps and $w=35$ for sticks, with $\Delta t=0.24$ ps. 
In general, $w$ must be chosen carefully.
If $w$ is too large, fluctuations that announce a predictor might be smoothed out and become undetectable.
Conversely, if $w$ is too small, there will be many fluctuations on different time scales in the predictor that are not relevant for events.
The values mentioned above were chosen because they produce the best ROC-curves.
%
%
\begin{figure*}[t!!!]
\vskip-2.3\bigskipamount
\centerline{
\hskip3\bigskipamount
\begin{minipage}[t!!!]{0.32\textwidth}
\epsfig{figure=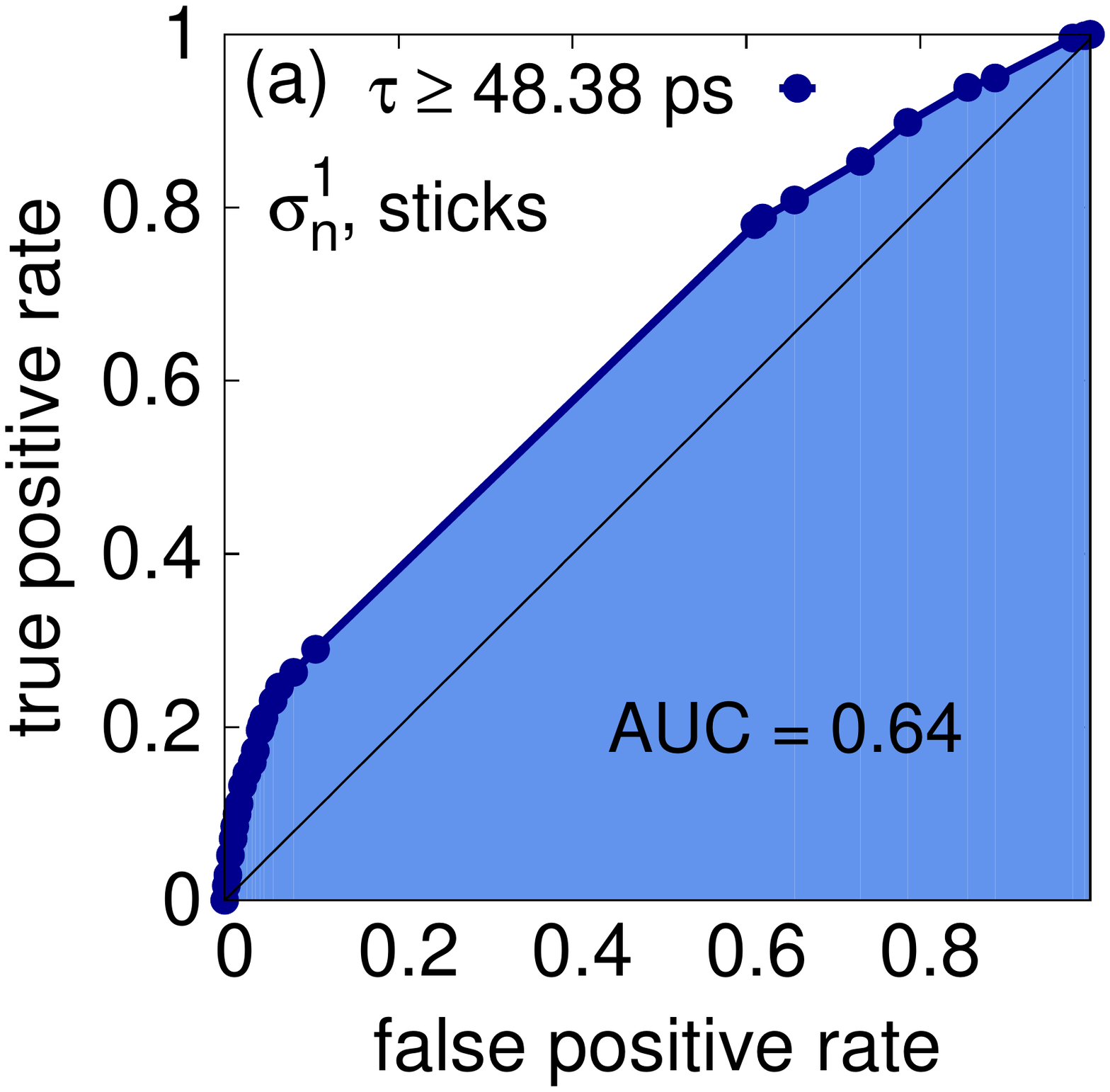,width=1.4\textwidth}
\end{minipage}
\hspace{-1.7cm}
\begin{minipage}[t!!!]{0.9\textwidth}
\epsfig{figure=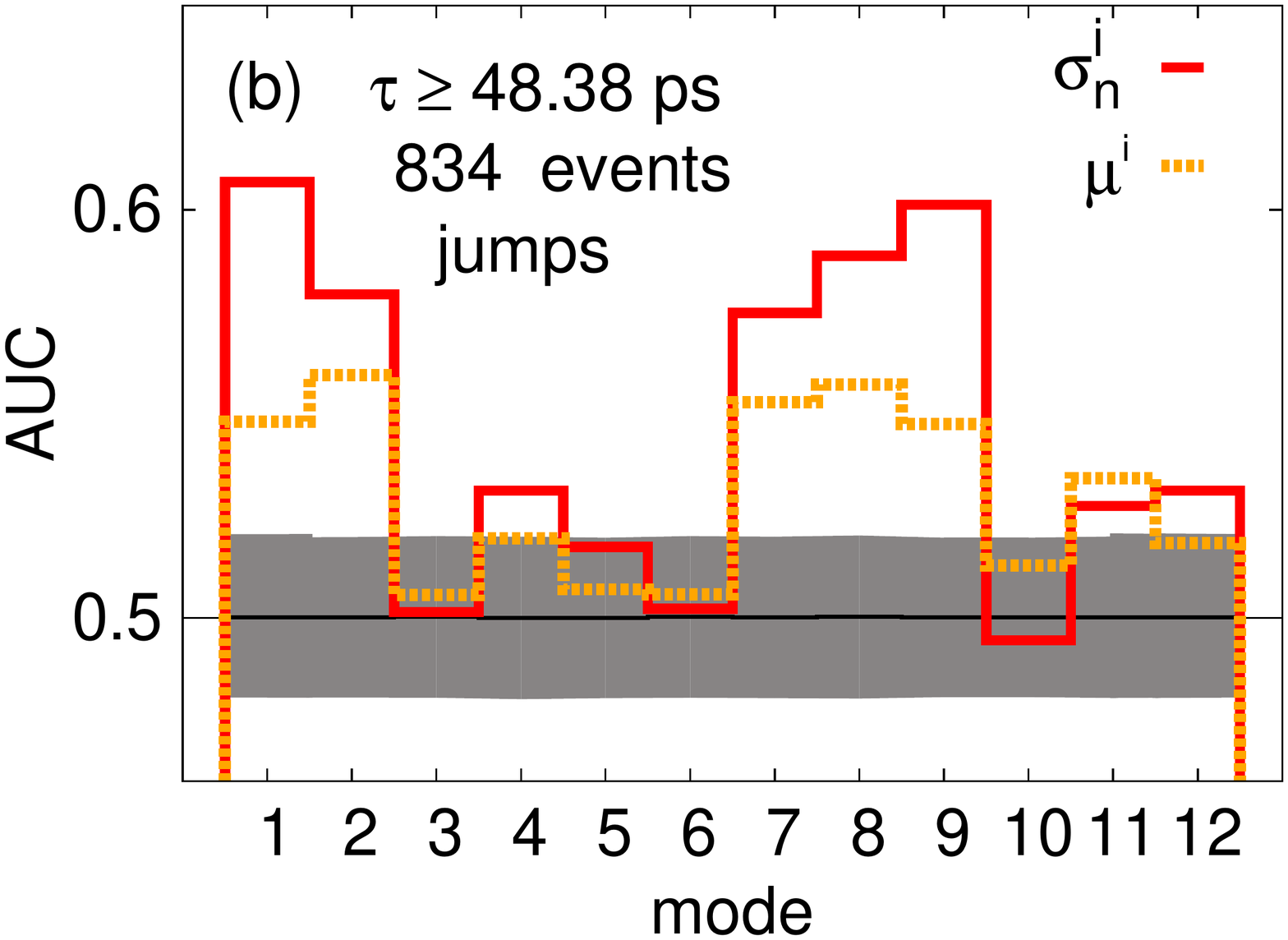,width=0.5\textwidth}
\hspace*{-2cm}
\epsfig{figure=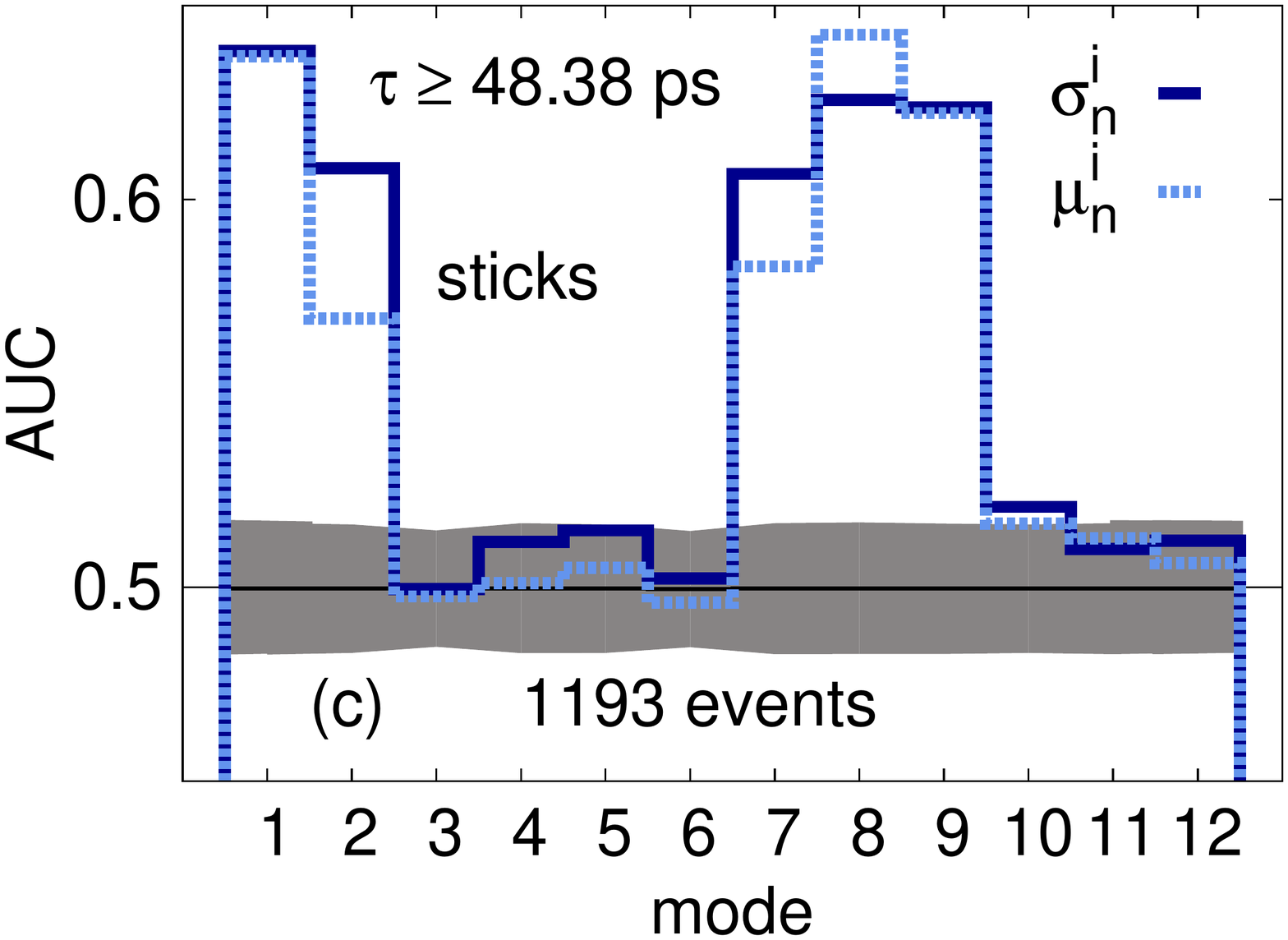,width=0.5\textwidth}
\end{minipage}
}
\vskip-0.5\bigskipamount
\caption{\label{fig:prec-compare}(Color online) Nowcasting ballistic jumps and subdiffusive sticks.
(a) An example for an ROC-curve.
(b) and (c): AUCs for different predictor variables $\mu^{i}_n$ and $\sigma^{i}_n$, estimated for {\sl nowcasts}, i.e., lead time $l=0$. The $95$\% confidence intervals are shown as shaded areas.
}
\end{figure*}

\begin{figure}[t!]
\vskip-1.8\bigskipamount
\hskip-1.2\bigskipamount
\begin{minipage}[t!!!]{0.5\textwidth}
\epsfig{figure=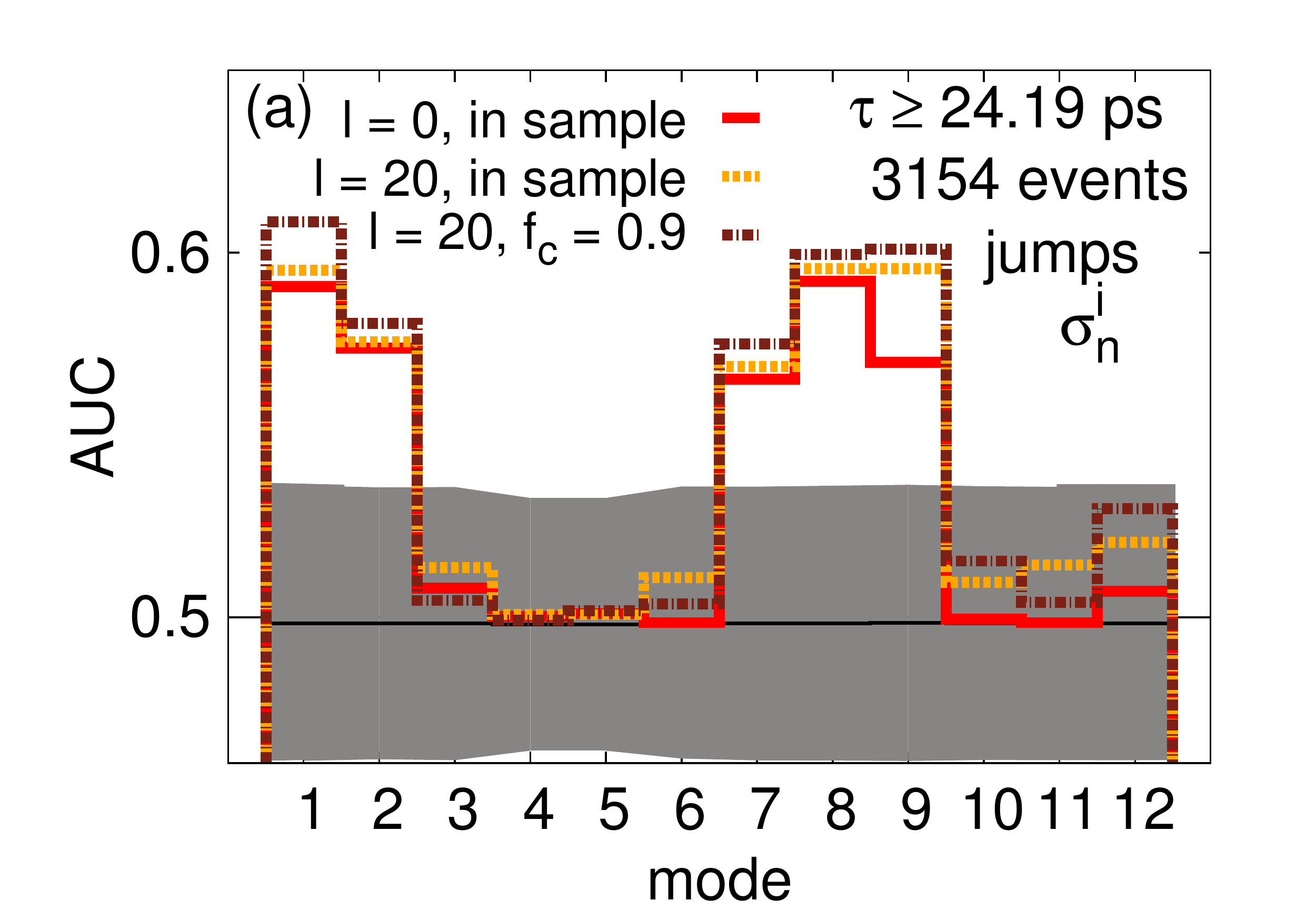,angle=-90,width=1.08\textwidth}\\
\vskip-0.8\bigskipamount
\epsfig{figure=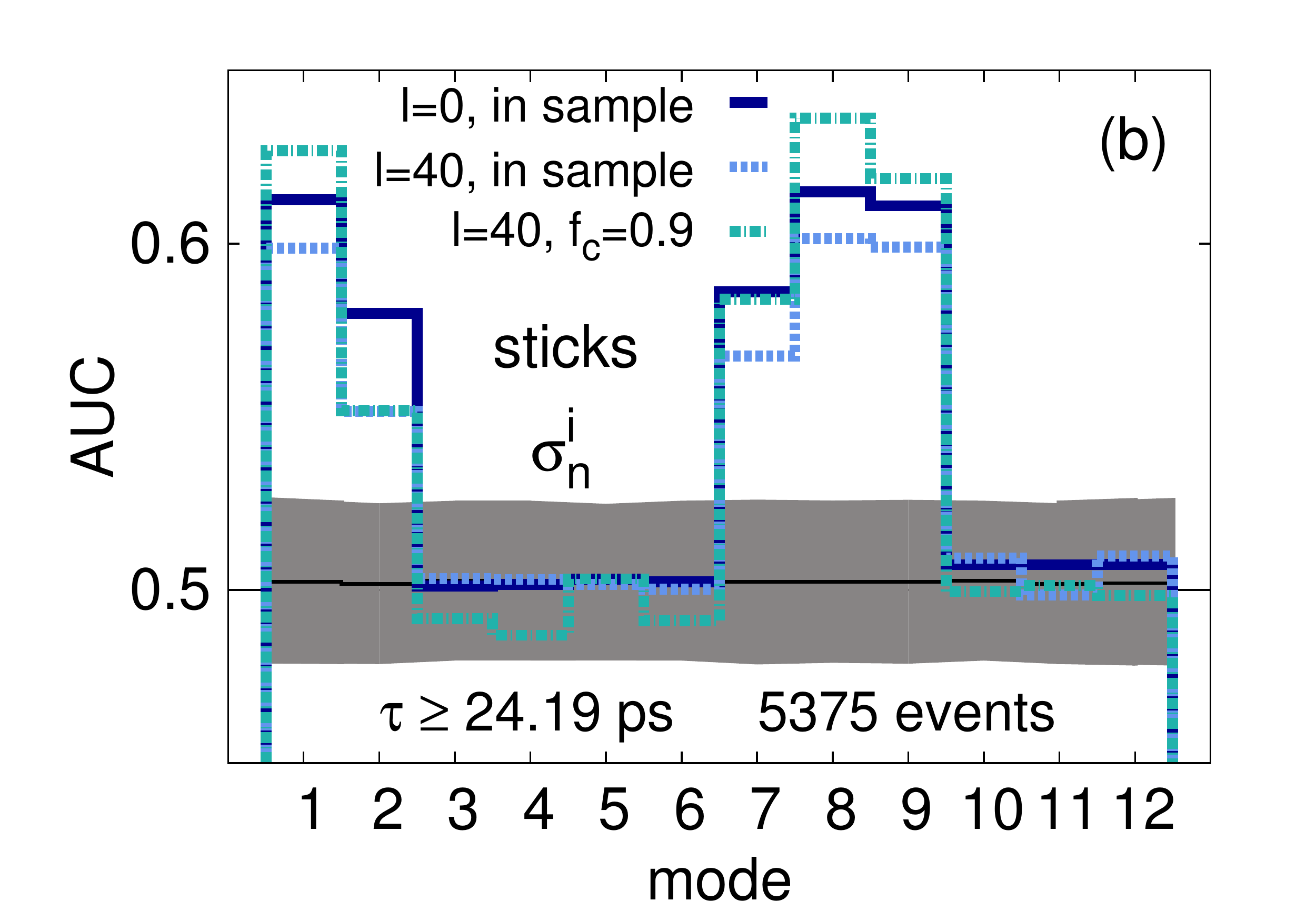,angle=-90,width=1.08\textwidth}
\end{minipage}
\vskip-0.5\bigskipamount
\caption{\label{fig:forecast-prec-compare}
(Color online) Forecasting long jumps and sticks.
(a) and (b): Comparison of nowcasts ($l=0$) and forecasts ($l>0$), both made using the standard deviation as a predictor. Additionally, we separated the data set into a test and a training part (90\% for training, 10\% for testing) which is indicated by the factor $f_c=9.0$. 
95\% Confidence intervals were also estimated through random predictions with parameters $l>0$ and $f_c=0.9$. 
}
\vskip-2\bigskipamount
\end{figure}

Relevant predictors are then identified using naive Bayesian classifiers~\cite{Rish2001}, i.e., conditional probability distribution functions (CPDFs) $p(\mbox{e}_{n+l} \geq \tau| y_n)$.
The event e$_{n+l}$ is either a long jump $j_{n+l}$ or stick $s_{n+l}$ starting at time instance $n+l$ in the future and lasting for a time $\tau \Delta t$ or longer.
The variable $l$ denotes the time difference between the time $n$ when the predictor $y_n$ was observed and the occurence of the event at time $n+l$.
In the context of (weather) forecasting $l$ is called {\sl lead time}. 
We study the connection between predictor variables and events for several values of l. 
Whereas investigating the now-cast szenario ($l=0$) emphasizes the link between predictors and events (as shown in Fig.~\ref{fig:forecast-prec-compare}), forecast szenarios ($l>0$, see Fig.~\ref{fig:prec-compare}) might be more relevant for applications.

The number of bins for each CPDF was adapted such that each bin has at least $2$ entries.

These links were revealed by comparing receiver operating characteristic-curves (ROC-curves) through their summary indices.
However, whether a predictor will be successful or not can often already by seen from the conditional probability distribution.
The black bars in Fig.~\ref{fig:cpdfs}  are the marginal distributions of both predictor variables, the sliding window average $\mu_{n,w}^{i}$ and the sliding window standard deviation $\sigma_{n,w}^{i}$.
The lines in Fig.~\ref{fig:cpdfs} show CPDFs $p(\mbox{e}_{n+l} \geq \tau| y_n)$ estimated for the event $\mbox{e}_{n+l}$ being either a ballistic flight or a stick, occurring at time $n+l$ and $y_n$ is one of the two predictors tested, namely $\mu_{n,w}^{i}$ or $\sigma_{n,w}^{i}$.

The most meaningful predictor, the value most likely to be followed by an event, is the one that maximizes the CPDF.
Relevant predictors should lead to nonflat CPDFs, such as displayed by full and dashed lines in Fig.~\ref{fig:cpdfs} (a), (b), (e), (f) in contrast to the flat CPDFs in Fig.~\ref{fig:cpdfs} (c), (d), (g), and (h).
A meaningful predictor should also be specific, i.e., not occur by chance without being related to an event. 
An indication of specificity is that the maximum of the CPDF does not coincide with a maximum of the marginal probability distribution function (PDF). 

In total we find qualitative differences between the distributions estimated using the energy in modes 1, 2, 7, 8, and~9 and the ones estimated based on modes 3, 4, 5, 6, 10, 11, and~12.
For both predictors $\mu^{i}_{n,w}$ and $\sigma^{i}_{n,w}$ and for both types of events, the CPDFs generated from time series of modes 1, 2, 7, 8, and~9 display structure, whereas the CPDFs obtained from modes 3, 4, 5, 6, 10, 11, and~12 are relatively flat.
Additionally, the marginal PDFs of modes 1, 2, 7, 8, and~9 possess several maxima, while the marginal PDFs of the other modes decay either slower as an exponential function or as a Gaussian.  

The marginal PDFs of modes 1, 2, 7, 8, and~9 also show a larger range of support, i.e., the average energy in these modes calculated from the linearized Hamiltonian is larger and so is the standard deviation in energy.
This is because in reality there are nonlinear terms in the energy, including nonlinear coupling terms between the modes, that also contribute to the energy.
The energy in these nonlinear terms can be comparable or larger than the energy in the linear terms.  If this were not the case, the system would not be so ubiquitously chaotic.
As we cannot assign the nonlinear mixing terms to specific degrees of freedom, it is impossible to calculate the energy in a specific mode more accurately, or indeed check the equipartition of thermal energy between the various modes.
The different nonlinear coupling of modes that are degenerate in the linearized system also leads to small quantitative differences in the PDFs between sets of degenerate modes.
%

%
Applying the CPDFs to make nowcasts and forecasts, we formulate a binary decision variable based on a probability threshold $\delta \in \left[0,\mbox{max} [p(e_{n+ l} \geq \tau | y_n)] \right]$ for each time step $n$
\begin{equation}
\label{eq:decision}
A_n = \left\{ \begin{array}{ll} 1 & \mbox{if } p(e_{n+ l} \geq \tau | y_n) \geq \delta,\\
0 & \mbox{otherwise}.  \end{array} \right. \quad 
\end{equation}
$A_n=1$ refers to issuing an alarm for an event to occur at time $n+l$, and $A_n =0$ to issuing no such warning.
%
%
%
The effectiveness of $A_n$ is evaluated by comparing the fraction of correct predictions out of all observed events (true positive rate) to the fraction of false alarms out of all non-events (false positive rate), i.e.,\ by generating ROC-curves~[see Fig.~\ref{fig:prec-compare}(a)].
Each value of the threshold $\delta$ corresponds to a single point in the ROC-curve.
An area under the curve (AUC) indicating better than random performance (curve on the diagonal) should have a value larger than $1/2$.
In order to estimate $95$\% confidence intervals for the AUCs, we additionally compute $100$~AUCs, generated by making random predictions.

As shown in Figs.~\ref{fig:prec-compare} and \ref{fig:forecast-prec-compare}, predictors based on modes 1, 2, 7, 8, and 9 have AUCs that are substantially higher than the $95$\% confidence intervals.
Similar results were obtained for longer and shorter event durations and when testing for the possibility of {\sl predicting} long-lived movements with a lead time $l > 0$ .
In this forecast scenarios [Fig.~\ref{fig:forecast-prec-compare}(a) and (b)] we separate between test and training data set, by estimating CPDFs on the first $f_c \cdot 100$\% of the data and generating ROCs and AUC on the remaining data.
Figure~\ref{fig:forecast-prec-compare}(a) and (b) indicate a certain forecast success for jumps and sticks up to $4.8$~ps before they occur, which is far in advance compared to the time scales of the internal dynamics of the molecule (about 0.5~ps).

Taking a closer look at the successful predictors as indicated by maxima of CPDFs, we find that a low standard deviation in modes 1, 2, 7, 8, and~9 can be associated with the occurrence of sticks, while a high standard deviation in these modes is observed simultaneously with the occurrence of long jumps.
Furthermore, the CPDFs of $\mu^{j}_{n}$ with $j=1,2,7,8$, and $9$, suggest that high values of the average energy can be associated with long jumps, whereas any deviation of $\mu^{j}_{n}$, with $j=1,2,7,8$, and $9$ from their most likely values can be associated with the occurrence of sticks. 
%

We can make an important observation about the modes that are connected to long-lived movements and the physical origin of this connection.
The modes with strong precursors display specific symmetries.
Mode 9 is a breathing mode symmetric under rotations by 60 degrees.
Mode 1, 2, 7, and 8 are mapped onto themselves by rotations over 180 degrees.
By contrast, the bending and stretching modes not showing predictors (3, 4, 5, and 6) are all antisymmetric under rotation over 180 degrees.
For antisymmetric vibrations, the coupling with the substrate with hexagonal symmetry is small or vanishes completely to leading order.
The torsion modes (10, 11, and 12) primarily involve motion in the $z$ direction,
and do not strongly couple to the motion in the $x$ and $y$ direction.
This is surprising, since there are clear links between the anomalous behavior and the torsional degrees of freedom.
Specifically, if torsion is removed completely, the diffusion of the model molecule is known to become anomalous~\cite{astridbenzene}.
However, as they do not couple directly to the center of mass, a small manipulation of the torsional degrees of freedom does not strongly affect the transport.
%

\section{Triggering long-lived movements}
\begin{figure}[ht!!]
\vskip-2.2\bigskipamount
\centerline{
\hskip1\bigskipamount
\epsfig{figure=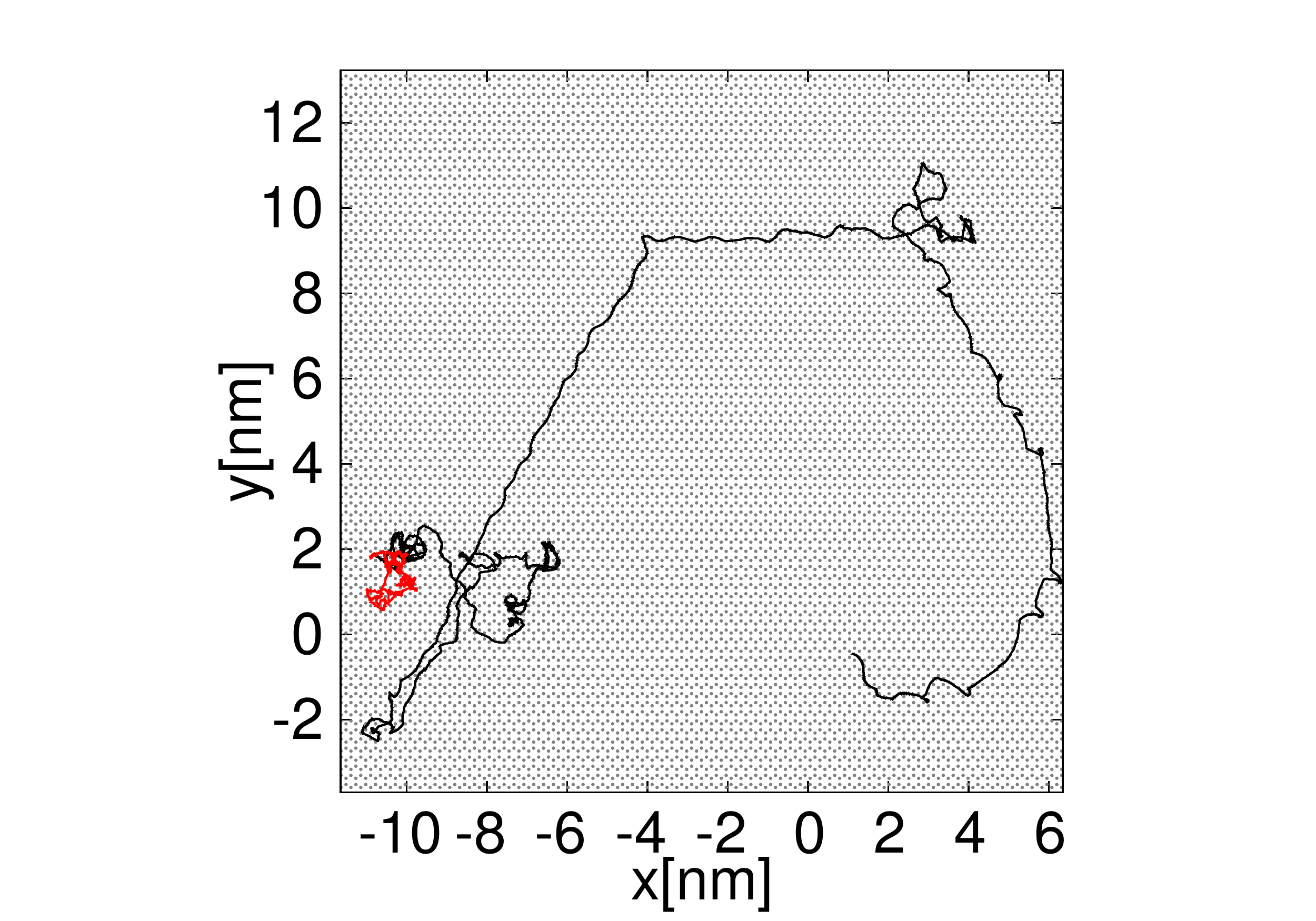, angle=-90, width=0.72\textwidth}
}
\vskip-0.5\bigskipamount
\caption{\label{fig:movie}(Color online) Damping mode one induces a long stick, i.e, the molecule remains within a region of the size of a few unit cells.
The trajectory is plotted in black without damping and in red (grey) after the damping started.
Note that both parts of the simulation represent the motion of the molecule during time intervals of equal length, i.e. 242 ps before the damping started and 242 ps with damping of mode one. 
This simulation is also included in form of a movie in the suplementary material
}
\end{figure}
\begin{figure}[ht!!]
\vskip-2.4\bigskipamount
\centerline{
\hskip1\bigskipamount
\epsfig{figure=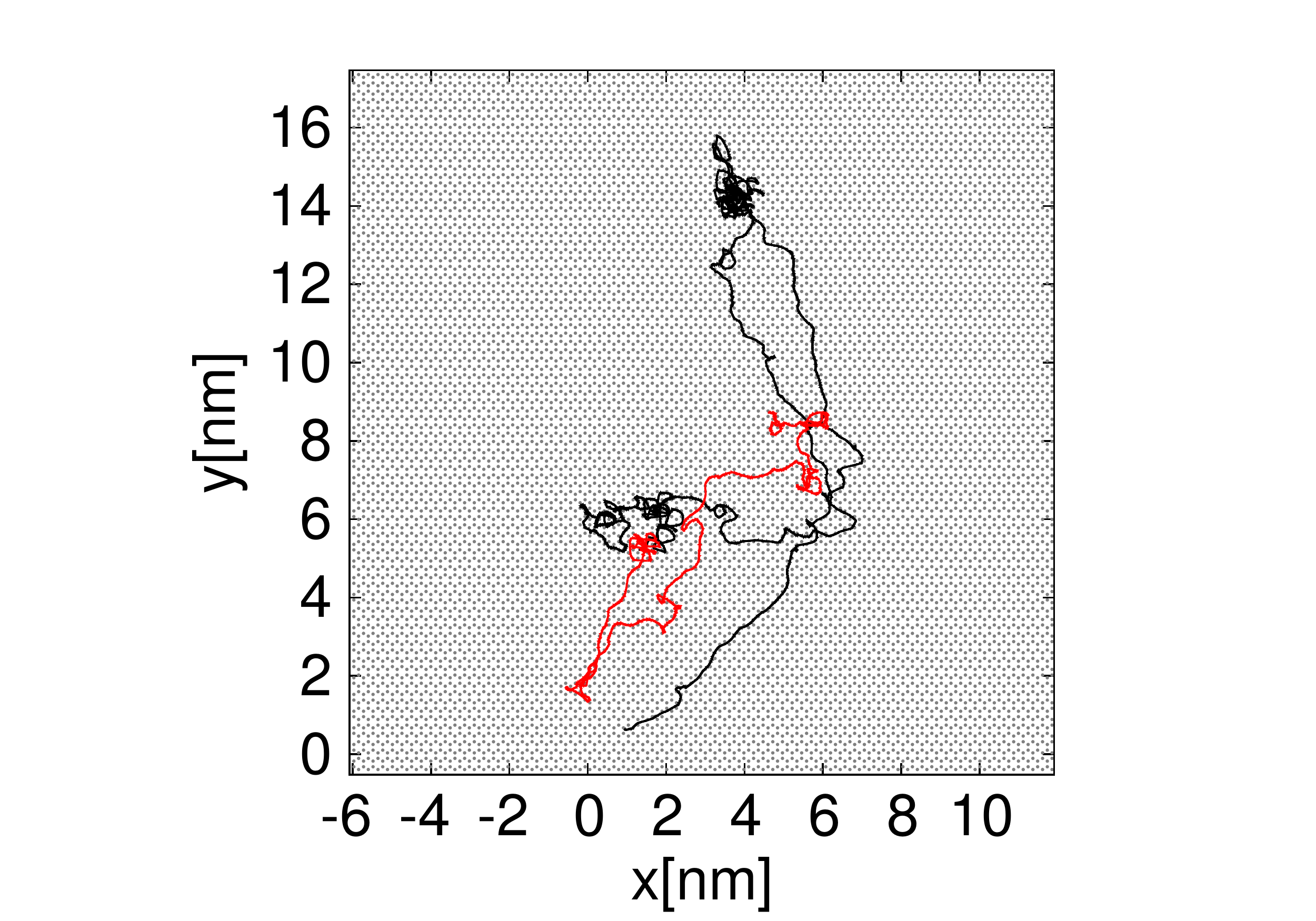, angle=-90, width=0.72\textwidth}
}
\vskip-0.5\bigskipamount
\caption{\label{fig:movie10}
(Color online) Damping of mode ten does not induce any qualitative change of the molecule's motion.
The molecule continues to diffuse over a wide range of the substrate, which is in contrast to the stick induced by the damping of mode one (see Fig.~\ref{fig:movie}).
}
\end{figure}
Having identified relevant predictors, one can trigger long-lived movements and thus manipulate diffusion. 
In an experiment, this could be accomplished by excitation of a specific vibrational mode with radiation. 
In our simulations, we achieve a similar effect by applying a viscous damping to a particular mode (see Fig.~\ref{fig:movie}).
In more detail we simulate the trajectory of a molecule for 242 ps without damping and then 242 ps with viscous damping of a particular mode.
Damping a relevant mode as e.g., mode one induces a stick, i.e., the molecule remains within a region of a few unit cells until the end of the simulation.
In contrast to this, damping of a non-relevant mode as e.g. mode ten has no apparent quantitative effect on the diffusion of the molecule (see Fig.~\ref{fig:movie10}).
We chose damping over driving the system because it suffices and keeps the system as simple as possible: Damping introduces only one extra parameter, the damping constant, rather than two, the frequency and amplitude of the driving.
Results are shown in Fig.~\ref{fig:prec-compare}.
\begin{figure}[t!]
\vskip-1.8\bigskipamount
\centerline{
\hspace{0.3cm}\epsfig{figure=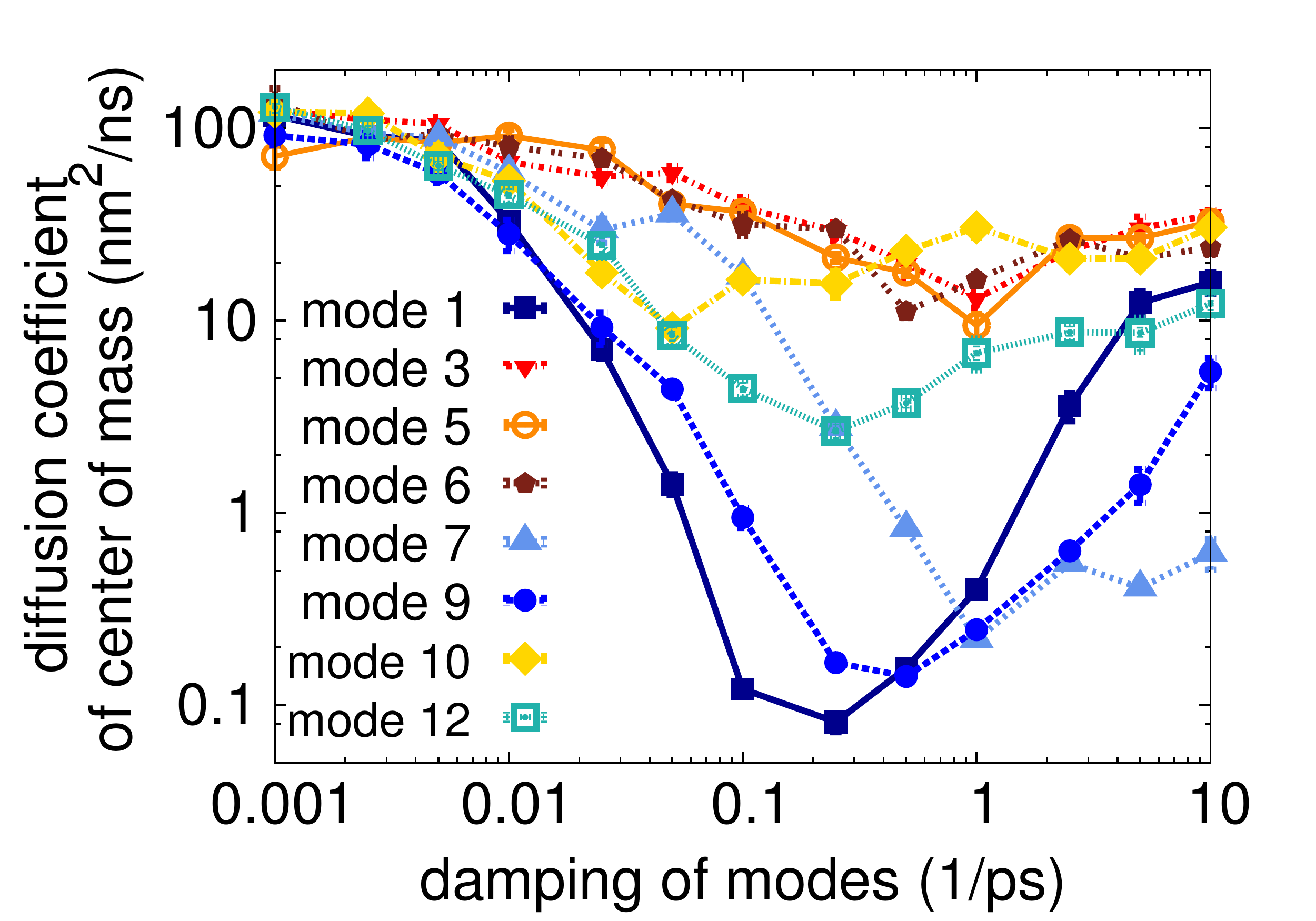, angle=-90, width=0.53\textwidth}
}
\vskip-0.5\bigskipamount
\caption{(Color online) Diffusion versus damping in a system with a single damped mode as indicated. 
For degenerate pairs, only one mode is shown.
There is a large drop in the diffusion coefficient when the modes with strong predictors are damped.}
\end{figure}
%
Diffusion decreases with damping for all modes, as the lower energy in the system makes it more difficult for the molecule to overcome the diffusion barrier.
However, for modes 1, 2, 7, 8, and~9,
we find an additional drop in the diffusion at relatively low damping, followed by a recovery. 
Note that these are exactly the modes providing relevant predictors with high AUC values.  
The recovery is likely related to the time scales of the dynamics of the center of mass on the substrate, which is around 1~ps.
When the damping is strong, the nonlinear dynamics of the center of mass on the substrate and in the internal degrees of freedom are changed qualitatively.  Consequently, jumps and sticks, if present at all, may no longer work in the same way.\\[0.5cm]

\section{Discussion}
In summary, we have demonstrated that long-lived jumps and sticks of complex molecules on substrates can be related to energies in {\it specific} internal degrees of freedom by using ROC analysis as a framework.
Apart from detecting links between the vibrational modes and simultaneously occurring long jumps or sticks, we have also studied the potential of this approach for predicting future long jumps and sticks.
In addition, we are able to deliberately trigger the long jumps and sticks, modifying the diffusion, by damping the modes that contain relevant predictors.

In nano-technological applications a manipulation method 
must be applicable to different molecules, and there must be
experimentally practical ways of implementing the control mechanism.
While for this proof of concept study we used a relatively simple prototype system,
our approach is applicable to larger, more complex, molecules, as it requires only a suffiently long phase-space trajectory generated by a molecular-dynamics simulation.
Moreover, the predictor variables can be chosen in any way that facilitates control in experimental settings.
Our results demonstrate that statistical inference has the potential to become a powerful method for studying high-dimensional dynamical systems in general, and understanding and manipulating molecular transport in particular.
%
%
\section{Acknowledgments}
 The authors are grateful to A.~Fasolino, E.~Altmann, W.~Just and G.~Radons for discussions.
 %
 ASdW has been financially supported by a Veni grant from the Netherlands Organization for Scientific Research (NWO) and by an Unga Forskare grant from the Swedish Research Council. 
 The collaboration of the two authors has additionally been supported by short visit grants from the European Science Foundation research networking program Exploring the Physics of Small Devices (EPSD).
\end{document}